\documentclass[useAMS,usenatbib]{mn2e}
\usepackage{color,graphicx} 
\usepackage{latexsym, times, setspace, amssymb, amsfonts, rotating} 
\usepackage[fleqn]{amsmath}
\usepackage[integrals]{wasysym}
\usepackage[tight]{subfigure}
\usepackage{natbib}

\usepackage{lscape}

\begin{document}

\title{Long-term Radio Observations of the Intermittent Pulsar
  B1931$+$24}
\date{\today}

\author[N.~J.~Young et al.]
{\parbox{\textwidth}{N.~J.~Young,$^{1,2}$\thanks{E-mail: young.neiljames@gmail.com} 
B.~W.~Stappers,$^1$ A.~G.~Lyne,$^1$ P.~Weltevrede,$^1$ M.~Kramer$^{1,3}$ and
I.~Cognard$^{4,5}$}\vspace{0.4cm}\\ 
\parbox{\textwidth}{$^1$Jodrell Bank Centre for Astrophysics, The
  University of Manchester, Alan-Turing Building, Manchester M13 9PL,
  United Kingdom \\ $^2$School of Physics, University of the
  Witwatersrand, Johannesburg Wits, 2050, South Africa
  \\$^3$Max-Planck-Institut f\"{u}r Radioastronomie, Auf dem H\"{u}gel
  69, 53121 Bonn, Germany \\ $^4$Laboratoire de Physique et Chimie de
  l'Environnement et de l'Espace, CNRS/Univ Orl\'{e}ans, F-45071
  Orl\'{e}ans, France\\ $^5$Station de radioastronomie de Nan\c{c}ay,
  Observatoire de Paris, CNRS/INSU, F-18330, Nan\c{c}ay, France}}
\maketitle
\begin{abstract}
We present an analysis of approximately 13-yr of observations of the
intermittent pulsar B1931$+$24 to further elucidate its behaviour. We
find that while the source exhibits a wide range of nulling
($\sim4-39$~d) and radio-emitting ($\sim1-19$~d) timescales, it cycles
between its different emission phases over an average timescale of
approximately 38~d, which is remarkably stable over many years. On
average, the neutron star is found to be radio-emitting for
$26\pm6\,\%$ of the time. No evidence is obtained to suggest that the
pulsar undergoes any systematic, intrinsic variations in pulse
intensity during the radio-emitting phases. In addition, we find no
evidence for any correlation between the length of consecutive
emission phases. An analysis of the rotational behaviour of the source
shows that it consistently assumes the same spin-down rates,
i.e. $\dot{\nu}=-16 \pm 1\times10^{-15}$~s$^{-2}$ when emitting and
$\dot{\nu}=-10.8 \pm 0.4\times10^{-15}$~s$^{-2}$ when not emitting,
over the entire observation span. Coupled with the stable switching
timescale, this implies that the pulsar retains a high degree of
magnetospheric memory, and stability, in spite of comparatively rapid
($\sim$~ms) dynamical plasma timescales. While this provides further
evidence to suggest that the behaviour of the neutron star is governed
by magnetospheric-state switching, the underlying trigger mechanism
remains illusive. This should be elucidated by future surveys with
next generation telescopes such as LOFAR, MeerKAT and the SKA, which
should detect similar sources and provide more clues to how their
radio emission is regulated.

\end{abstract}

\section{Introduction}\label{sec:intro}
PSR~B1931$+$24 was discovered in 1985 with the NRAO 100-m Green Bank
Telescope \citep{stwd85}. However, it was not until 13 years later,
during routine pulsar timing observations at Jodrell Bank, that the
remarkable properties of this object were noted. Unlike conventional
nulling pulsars, which undergo temporary emission cessation (or
\emph{pulse nulling}) over timescales of $\lesssim100$ pulse periods
(e.g. \citealt{bac70,rit76,ran86,wmj07}), PSR~B1931$+$24 has been
found to exhibit extremely long duration, active ($\sim5-10$~d) and
quiescent ($\sim25-35$~d) radio emission phases. Remarkably, these
emission phases also repeat quasi-periodically, are broadband in
frequency and are found to be correlated with the rotational behaviour
of the star. That is, the source exhibits a spin-down rate which is
$\sim50\,\%$ greater in the active (radio-on, hereafter) emission
phases, compared with that of the non-radio emitting (radio-off,
hereafter) phases \citep{klo+06}.

To date, a handful of similar objects have been discovered,
e.g. PSR~J1832$+$0029 \citep{llm+12} and PSR~J1841$-$0500
\citep{crc+12}. Despite this, only the properties of PSR~B1931$+$24
have been studied, or attempted to be explained, in some detail (see,
e.g., \citealt{klo+06,cs08,rks+08,rmt11,lst12a}). As such, theories
which attempt to explain the behaviour of such long-term intermittent
pulsars are centred mainly on observations of this object,
particularly those in the radio regime\footnote{While PSR~B1931$+$24
  has been observed both in the radio \citep{stwd85,klo+06} and
  optical \citep{rks+08} regimes, it has so far only been detected in
  radio observations.}.

Amongst the most promising of these theories, is the possibility that
the pulsar magnetosphere experiences systematic variation in its
global charge distribution; that is, it undergoes
`magnetospheric-state switching'
(e.g. \citealt{bmsh82,con05,lhk+10,tim10,lst12a}). In this context, it
is proposed that global re-distributions of current are responsible
for causing alterations to the morphology of the emitting region,
which result in the presence (or absence) of radio emission, as well
as fluctuations in the spin-down rate via associated changes in the
magnetic field configuration. However, it is unclear how these
alterations occur, nor what their intrinsic timescales should be; an
array of vastly different trigger mechanisms have been proposed,
e.g. orbital companions \citep{rks+08,cs08}, non-radial oscillations
\citep{rmt11}, precessional torques \citep{jon12}, magnetic field
instabilities \citep{grg03,ug04,rkg04} and polar cap surface
temperature variations \citep{zqlh97}, but no consensus has been
reached on a definitive mechanism. Subsequently, little is known about
the processes which govern the behaviour of intermittent pulsars, nor
what their relationship is with conventional nulling pulsars and
rotating radio transients (RRATs; \citealt{mll+06}).

In this paper, we use an unparallelled span of observations to probe
the emission and rotational properties of PSR~B1931$+$24 in detail. We
focus on expanding upon the results of \cite{klo+06}, by using our
longer observation baseline to investigate both short- and long-term
variations in the behaviour of the pulsar. We present the details of
the radio observations in the following section. In
Section~\ref{sec:mod}, we describe the emission variability of the
source. This is followed by an investigation of its rotational
stability in Section~\ref{sec:spin}. We discuss the main implications
of our findings in Section~\ref{sec:discuss} and present our
conclusions in Section~\ref{sec:conc}.

\section{Observations}\label{sec:obs}
In this work, we present an approximately 13-yr span of observations
(29~April~1998 to 19~May~2011), including previously published data
(see \citealt{klo+06}), which has been used to investigate the
emission and rotational properties of PSR~B1931$+$24. These data were
predominantly obtained with the \hbox{76-m}~Lovell Telescope and the
28$\times$25-m~Mark~II Telescope at Jodrell Bank. Data was also
obtained with the 94-m~Nan\c{c}ay Radio Telescope in France, so as to
bridge a gap in observations during a period of extended telescope
maintenance\footnote{Between 20~October~2004 and 2~February~2005, the
  Lovell Telescope was unable to observe PSR~B1931$+$24 due to the
  discovery and replacement of a cracked tire. The Mark~II Telescope
  was used for continued timing measurements of other sources during
  this time.}. Two back-ends were used to acquire the Lovell
observations: the Analogue Filter Bank (AFB; up~to~May~2010) and the
Digital Filter Bank (DFB; since~January~2009). Table~\ref{tab:obs}
shows the typical observing characteristics for these instruments and
the Nan\c{c}ay data. We note that while the average observing cadence
is less than once per day for the AFB, a series of more intense
observations, i.e.  approximately twice-daily monitoring, was
initiated from 2006 onwards to provide better constraints on the
emission phase transition times.

\begin{table}
\caption{System characteristics for observations of PSR~B1931$+$24
  from 29~April~1998 to 19~May~2011. The total time span and number of
  observations are denoted by $T$ and $N$ respectively. The most
  common observation duration is represented by $\Delta t$ and the
  typical cadence is denoted by $C$. The typical centre frequency,
  bandwidth and channel bandwidth of the observations are given by
  $\nu$, $B$ and $B_{\mathrm{ch}}$ respectively.}  
\centering
\begin{tabular}{r r r r r}
  \hline
  \hline
  System Property          & AFB  & DFB & Mark~II & Nan\c{c}ay \\ \hline 
  $T$ (d)\dotfill          & 4403 & 856  & 3661 & 243  \\
  $N$\dotfill              & 3996 & 1168 & 636  & 78   \\
  $\Delta t$ (min)\dotfill & 12   & 12   & 42   & 13   \\
  $C$ (d$^{-1}$)\dotfill   & 0.9  & 1.4  & 0.2  & 0.3  \\
  $\nu$ (MHz)\dotfill      & 1402 & 1520 & 1396 & 1368 \\
  $B$ (MHz)\dotfill        & 32   & 384  & 32 & 64     \\
  $B_{\mathrm{ch}}$ (MHz)\dotfill & 1    & 0.5  & 1  & 4      \\
  \hline
\end{tabular}
\label{tab:obs}
\end{table}

\section{Emission Variability}\label{sec:mod}
\subsection{Pulse intensity modulation}
\begin{figure*}
  \begin{center}
    \includegraphics[trim= 20mm 20mm 5mm 10mm,clip,angle=270,totalheight=15in,height=10.2cm,width=18.5cm]{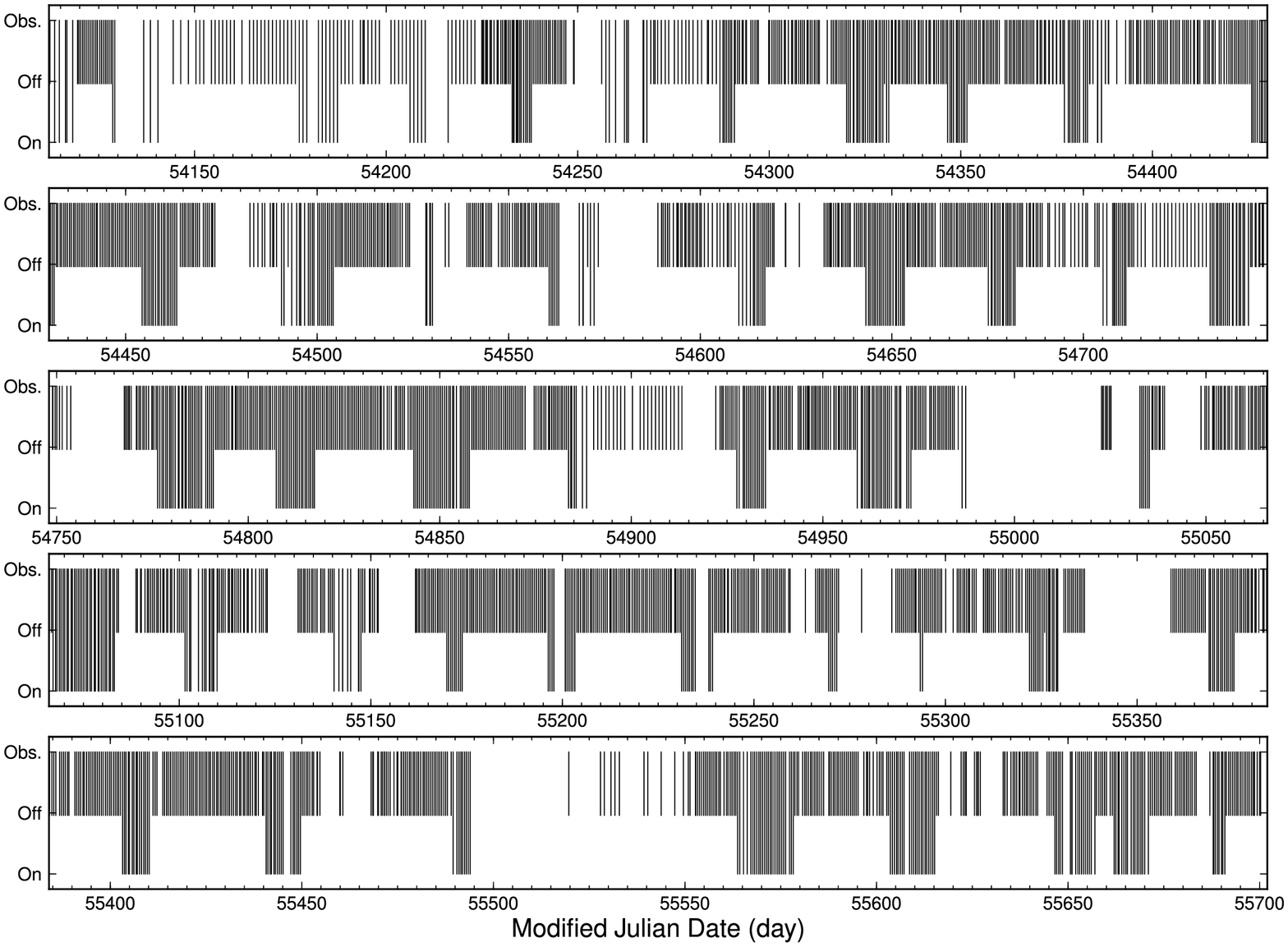}
   \end{center}
 \vspace{-5pt}
\caption{Sequence of observations, denoted by the black lines, carried
  out over $\sim4$~yr (12~January~2007 to 19~May~2011). The data is
  separated into 5 contiguous 318-d panels. The times of observation
  and the times when PSR~B1931$+$24 was radio-on (full height) and
  radio-off (half height) are shown by the extent of these lines.}
\label{onoff}
\end{figure*}

Following the work of \cite{klo+06}, we sought to elucidate the
emission variability in PSR~B1931$+$24, based on a longer observation
baseline and higher cadence data. In particular, we were interested in
further characterising the timescales of emission variation, and the
durations of the radio-on and -off emission phases. For the purpose of
this work, we used the activity duty cycle (ADC) method discussed in
\cite{ysw+12}. Here, we visually inspected the average pulse profile
data, which was formed over the entire integration length and
bandwidth of observations, and constructed a time-series of one-bit
data corresponding to the `radio activity' of the pulsar; that is, 1's
for profiles consistent with the emission signature of the source and
0's for noise dominated profiles (i.e. radio-off states). The result
of this analysis for the highest cadence observations (i.e. the last
$\sim4$~yr of data) is shown in Fig.~\ref{onoff}.

\begin{figure*}
  \begin{center}
    \includegraphics[trim = 5mm 23mm 0mm 28mm,clip,height=16cm,width=6.8cm,angle=270]{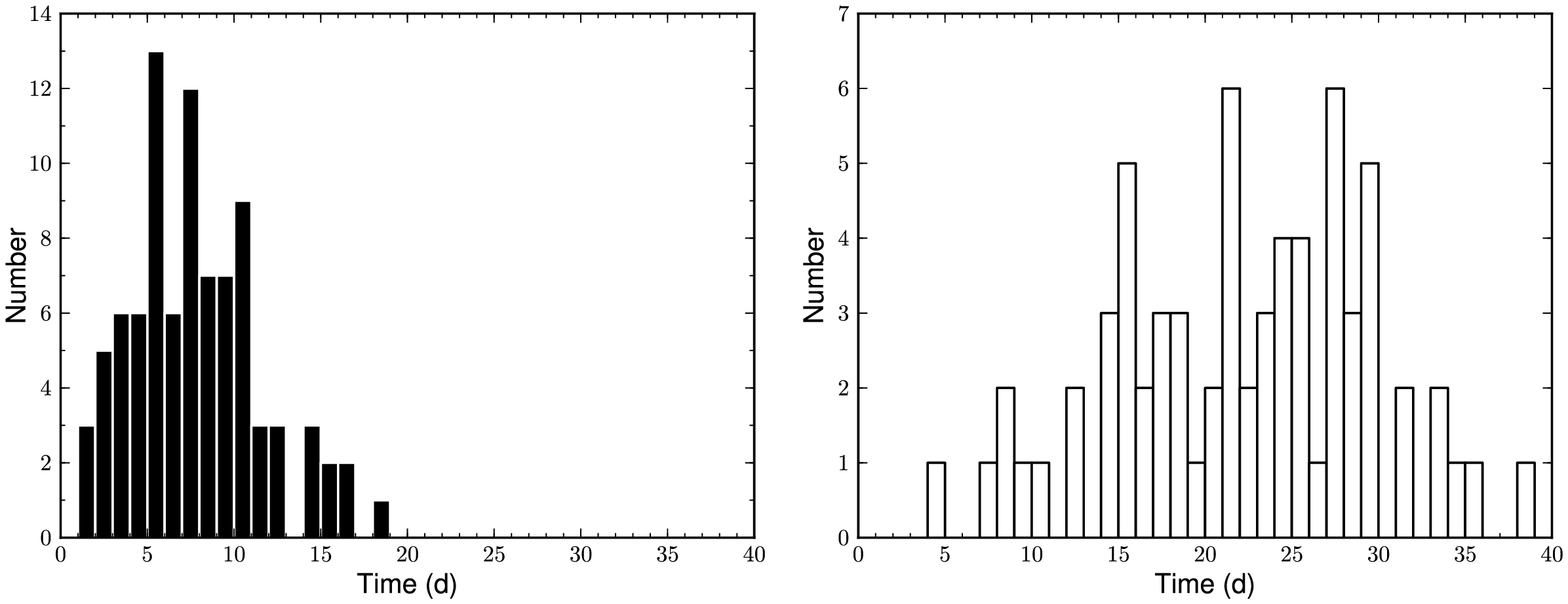}
  \end{center}
 \vspace{-5pt}
\caption{Histograms showing the durations when PSR~B1931$+$24 is
  observed to be in a particular emission phase, radio-on
  (\emph{left}) and radio-off (\emph{right}).}
\label{hists}
\end{figure*}

In the data set, some of the observations which define the timescales
of emission are separated by several days, particularly at earlier
epochs. Therefore, a number of the inferred emission phase durations
in the data do not accurately represent the behaviour of the
pulsar. Consequently, we only consider emission phase durations which
have gaps $\lesssim5$~d between consecutive observations, so that we
can pinpoint transition times (taken to be the midpoints between
radio-on and -off observations) as accurately as possible. After
applying these criteria we were left with only the highest confidence
emission phase durations, which are shown in Fig.~\ref{hists}. We note
that there is no evidence for emission cessation over timescales less
than a day. The average times in each emission phase are \hbox{$8 \pm
  4$~d} and \hbox{$22 \pm 7$~d} for the radio-on and -off phases
respectively. These values are consistent with the average radio-on
and -off timescales observed in \cite{klo+06}, that is $\sim6$~d and
$\sim28$~d respectively. We note here that the radio-on and -off
timescales have a wider range of values than previously thought
(Fig.~\ref{hists}, c.f. Fig.~1c of \citealt{klo+06}). The improvement
in the determination of these values is due to the increased data span
and observation cadence.

We used these same emission phase durations to calculate the ADC; that
is, the percentage of time spent in the radio-on phase, compared with
the total observation time. Given that there are gaps between
consecutive emission phases, and that there is a disparity between the
number of suitable radio-on and -off phases, we determined the average
total time spent in each emission phase through bootstrapping the
observed distributions of the emission phase durations. That is, the
data was sampled with replacement to obtain $10^6$ resamples, and
average durations, for each distribution of emission phases. The
resultant values were used to calculate the average ADC over the
entire data-set. The error in the ADC was calculated from the
uncertainty in the total time spent in the radio-on phase, which was
obtained from bootstrapping the observed uncertainties. This results
in an $\mathrm{ADC}= 26 \, \pm6\,\%$ which is larger than, but
consistent with, the \cite{klo+06} value of $19\pm5\,\%$. We note that
the larger uncertainty in our value most likely results from slight
changes in observation cadence throughout the $\sim13$-yr data set.

\begin{figure}
  \begin{center}
    \includegraphics[trim = 5mm 2mm 4mm 5mm, clip, height=8.3cm,width=6cm,angle=270]{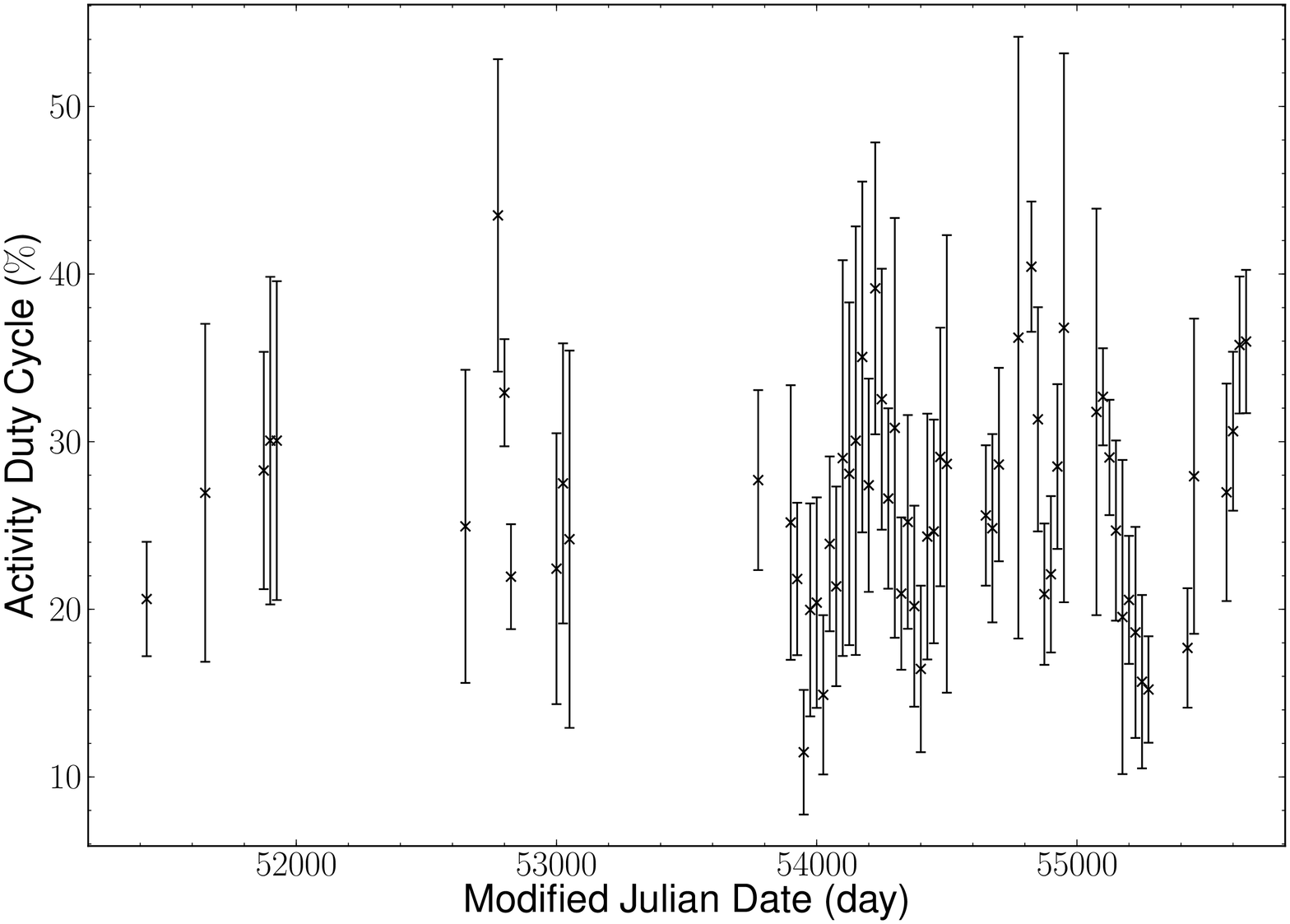}
  \end{center}
 \vspace{-5pt}
\caption{The evolution of the activity duty cycle of PSR~B1931$+$24
  over approximately 13~yr of data.}
\label{ADC}
\end{figure}

As our average ADC is the same as in \cite{klo+06} and yet the range
of radio-on and -off durations is wider, we wanted to ascertain if
there was any temporal evolution in these properties. Following the
stride fitting method of \cite{lhk+10}, we determined the ADC over
segments of length $T=100$~d\footnote{The interval length $T$ was
  chosen to provide a compromise between resolution in the ADC and our
  sensitivity to short-term noise variations, that is short-term
  fluctuations which do not reflect the typical behaviour of the
  object.}, offset by intervals of $T/4=25$~d across the data set. We
initially estimate the uncertainty in ADC for each data segment using
the sum of the errors in the transition times between consecutive
emission phase durations. Due to the irregular time sampling, however,
there are also gaps between consecutive observations during emission
phases. As the minimum time spent in a consecutive radio-on and
\hbox{-off} phase is approximately $10$~d, we assume that any gap
between a neighbouring observation which exceeds this duration should
contribute to the total uncertainty in the ADC. In order to compromise
between data accuracy and volume, we apply cut-offs to the
uncertainties in the data; that is, we do not consider intervals which
contain observations separated by more than $25$~d or whose error in
the ADC is greater than $50~\%$. The resultant ADC as a function of
time is shown in Fig.~\ref{ADC}.

We average these values (Fig.~\ref{ADC}) to obtain \hbox{$\langle
  \mathrm{ADC}\rangle = 26 \pm7\%$}, which is consistent with that
obtained by bootstrapping the emission durations. We attribute the
significant error in this quantity to the notable fractional
uncertainties in the ADC throughout the data set. By implementing an
Anderson-Darling test\footnote{This test is adapted from the
  Kolmogorov-Smirnov test to have greater sensitivity towards the
  tails of a distribution \citep{ptvf92}, which make it better suited
  to testing for normality.}  \citep{ptvf92} we find that the ADC
values are consistent with coming from a normal distribution, assuming
Gaussian statistics for the uncertainties. There is, therefore, no
compelling evidence for significant variation in the ADC of
PSR~B1931$+$24.

\begin{figure}
  \begin{center}
    \includegraphics[trim = 1mm 1mm 2mm 5mm, clip, height=8.2cm,width=8.2cm,angle=270]{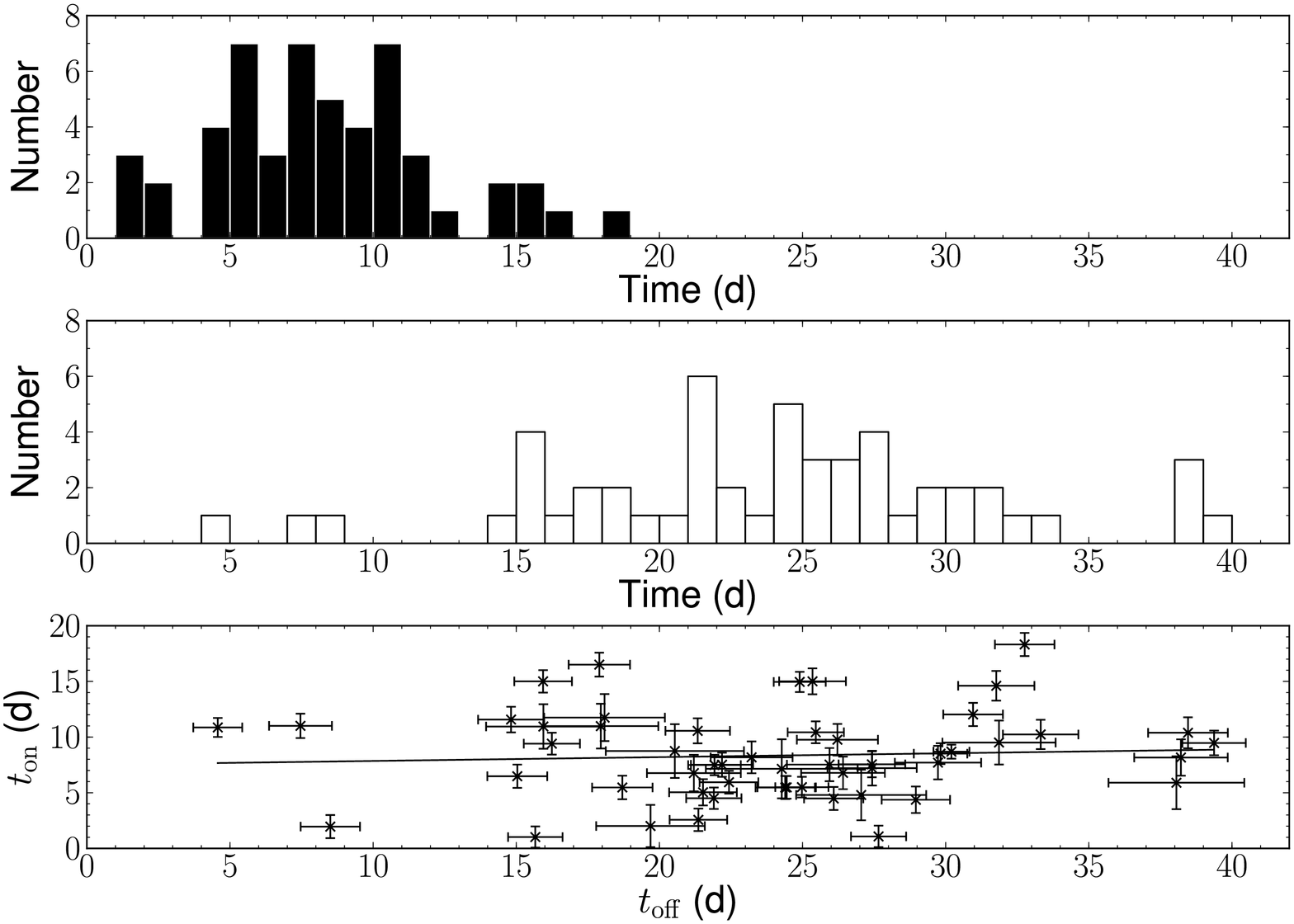}
  \end{center}
 \vspace{-5pt}
\caption{The correlation between the radio-on and -off emission phase
  durations of PSR~B1931$+$24. Emission duration histograms for all
  the radio-on \hbox{(\emph{top panel})} and radio-off
  \hbox{(\emph{middle panel})} phases from the observation periods
  defined in Table~\ref{tab:onoff_corr}, as well as the linear
  regression of the two parameters \hbox{(\emph{bottom panel})}. There
  is no evidence for a correlation between consecutive radio-on and
  radio-off interval lengths.}
\label{onoff_corr}
\end{figure}

We also sought to determine whether there were any interdependencies
in the characteristics of the pulsar emission. In particular, we were
interested in discovering whether there is a correlation between the
length of successive radio-on and -off phases. For this analysis, we
selected nine data intervals from the entire data set which had the
highest observation cadence and, thus, the highest confidence in
emission phase durations. The properties of these data intervals are
listed in Table~\ref{tab:onoff_corr}. We correlated the durations of
successive radio-on and -off emission phases, for each of these data
intervals, to determine whether there was any connection between the
length of consecutive emission phases (see Fig.~\ref{onoff_corr}).

\begin{table*}
\caption{Properties of the nine data intervals chosen to characterise
  the correlation between the radio-on and -off emission phase
  durations of PSR~B1931$+$24. The start and finish times of each data
  interval, of length $T$, are denoted by MJD$_{\mathrm{start}}$ and
  MJD$_{\mathrm{finish}}$ respectively. The number of emission phases
  which are radio-on and -off are given by $N_{\mathrm{on}}$ and
  $N_{\mathrm{off}}$ respectively. The average total time spent in a
  radio-on and -off phase are $\bar{t}_{\mathrm{on}}$ and
  $\bar{t}_{\mathrm{off}}$, and the activity duty cycle of the pulsar
  for each data interval is denoted by ADC.}
 \label{tab:onoff_corr} 
\centering
\vspace{4pt}
\begin{tabular}{ r  r  r  r  r  r  r  r }
  \hline
  \hline
  MJD$_{\mathrm{start}}$ & MJD$_{\mathrm{finish}}$ & $T$ (d) & $N_{\mathrm{on}}$ & $N_{\mathrm{off}}$ & $\bar{t}_{\mathrm{on}}$ (d) & $\bar{t}_{\mathrm{off}}$ (d) & ADC ($\%$) \\
  \hline
  51814.07 & 51924.38 & 110.31 & 4 & 3 & $7\pm 1$ & $24\pm 2$ & $22\pm 8$ \\
  52762.52 & 52921.53 & 159.01 & 5 & 4 & $10\pm 4$ & $27\pm 4$ & $28\pm 2$ \\
  53733.23 & 53832.01 & 98.78  & 4 & 3 & $7\pm 2$ & $22\pm 7$ & $29\pm 5$ \\
  53889.35 & 54084.26 & 194.91 & 8 & 7 & $5\pm 1$ & $21\pm 3$ & $19\pm 5$ \\
  54176.30 & 54504.73 & 328.43 & 11 & 10 & $9\pm 1$ & $23\pm 2$ & $27\pm 5$ \\
  54609.06 & 54743.66 & 134.60 & 5 & 4 & $8\pm 1$ & $23\pm 1$ & $26\pm 4$ \\
  54775.96 & 54973.34 & 197.38 & 6 & 5 & $11\pm 2$ & $26\pm 4$ & $30\pm 4$ \\
  55066.68 & 55489.50 & 422.82 & 12 & 12 & $7\pm 1$ & $28\pm 2$ & $21\pm 3$  \\
  55563.29 & 55688.01 & 124.72 & 4 & 4 & $12\pm 1$ & $19\pm 6$ & $38\pm 3$ \\
  \hline
\end{tabular}
\end{table*}

We find that there is no significant correlation between the length of
time in a given emission phase and that of the opposing mode
consecutive to it. This suggests that the pulsar does not retain a
memory of the length of its previous magnetospheric-state after a
transition. We also computed the ADC for each data interval to
determine whether the results of the previous analysis may have been
biased by observation cadence (see Table~\ref{tab:onoff_corr}). We
again find that there is no significant evidence for variation in the
ADC over time.

\begin{figure}
  \begin{center}
    \includegraphics[trim = 3mm 2mm 2mm 5mm, clip,height=8.3cm,width=5.5cm,angle=270]{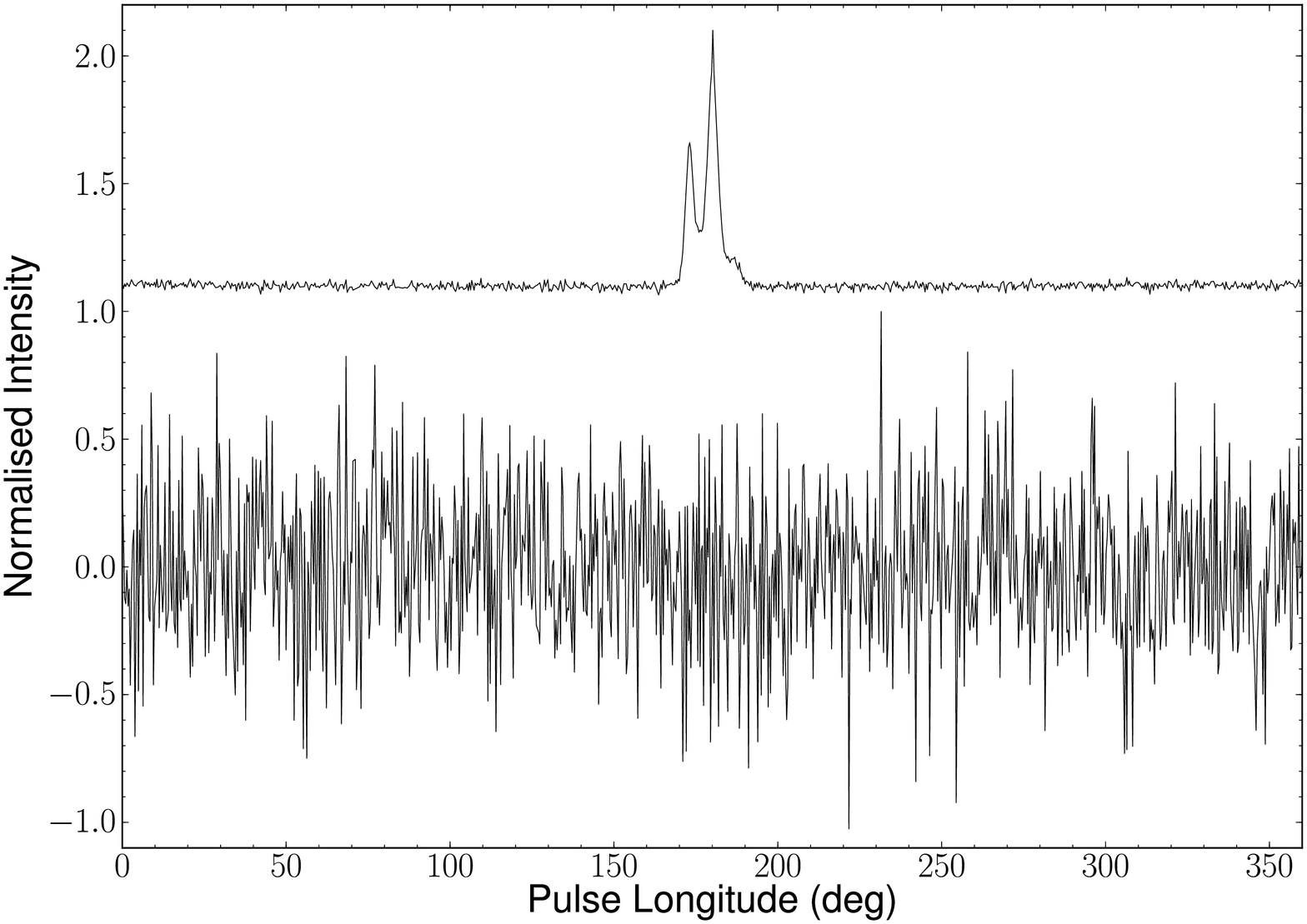}
  \end{center}
 \vspace{-5pt}
\caption{Average pulse profiles of PSR~B1931$+$24 showing the
  characteristics of the radio-on (\emph{top}) and radio-off emission
  (\emph{bottom}). The peak pulse intensities are normalised to one
  for each emission mode profile. The radio-on profile is offset in
  intensity from the radio-off profile for clarity.}
\label{limits}
\end{figure}

We were also interested in determining the average flux density of the
pulsar, attributed to the different phases of emission, using the
modified radiometer equation \citep{lk05}:
\begin{equation}
S = \frac{\beta \,\textrm{SNR} \,T_{\mathrm{sys}}}{G\sqrt{n_{\mathrm{p}}\,B\,T}}\sqrt{\frac{W_{\mathrm{eq}}}{P-W_{\mathrm{eq}}}}\,.
\label{eq:flux}
\end{equation}
Here, $\beta\sim1$ is the digitisation factor, $G\sim1$~Jy~K$^{-1}$ is
the telescope gain, $T_{\mathrm{sys}}\sim35$~K is the average system
temperature, $n_{\mathrm{p}}=2$ is the number of polarisations,
$B=384$~MHz is the observing bandwidth, $P=814$~ms is the pulsar
period and \hbox{$W_{\mathrm{eq}}=14$~ms} is the equivalent pulse
width. In total, we averaged 211 radio-on and 106 radio-off DFB
observations (see Fig.~\ref{limits}), to obtain total integration
times ($T$) of approximately 40~hr and 20~hr
respectively. Consequently, we place a limit on the mean flux density
in the radio-off phase $S_{\mathrm{off}}\lesssim2.0\pm0.4~\mu$Jy
(assuming a limiting $\mathrm{SNR}\sim3$). For the radio-on phase, we
obtain a mean flux density $S_{\mathrm{on}}=40\pm8~\mu$Jy
($\mathrm{SNR}\sim90$), which is at least 20 times brighter than that
in the radio-off phase.

Converting the flux densities into pseudo-luminosities, using
$L_{1400}\equiv S_{1400}\,d^2$ \citep{lk05}, we find
$L_{1400,\,\mathrm{off}}\lesssim0.04$~mJy~kpc$^2$ and
$L_{1400,\,\mathrm{on}}\sim0.84$~mJy~kpc$^2$ (for a pulsar distance
$d\sim4.6$~kpc, from the NE2001 model; \citealt{cl02}). We note that
the upper limit on the pseudo-luminosity of PSR~B1931$+$24 in the
radio-off phase is barely brighter than the two weakest known radio
pulsars i.e. PSR~J0030+0451 ($L_{1400}\sim0.04$~mJy~kpc$^2$;
\citealt{lzb+00}) and J2144$-$3933 ($L_{1400}\sim0.02$~mJy~kpc$^2$;
\citealt{lor94}). This implies that the radio-off phases are
consistent with emission cessation. However, it may be that the source
exhibits extremely weak, underlying emission which is far below our
detection threshold. This could be compared to PSR~B0823$+$26, another
intermittent radio source, which exhibits a factor of 100 difference
in intensity between its separate emission modes (\citealt{ysw+12},
c.f. \citealt{elg+05}).

Observations of PSR~B1931$+$24 by \cite{klo+06} suggest that it
exhibits enhanced particle flow during its radio-on phases. We were
interested in seeing whether other, less severe, changes in particle
flow occurred which might also be attributed to the mechanism which
determines its emission behaviour. One way to test whether the
emission in these phases was consistent with a `steady-state' particle
flow (c.f. Crab enhanced emission, see \citealt{sso+03}), was to
determine whether the pulsar exhibited any systematic pulse intensity
fluctuations during radio-on phases. Such analysis is ideally
performed using single-pulse data which, in this instance, was not
available. Therefore, we used the 12-min average pulse profiles to
determine whether there were any variations in pulse intensity, which
might correlate with the position of an observation around an emission
phase transition (c.f. pulse intensity decay and rise times around
nulls in PSR~B0809$+$74; \citealt{la83,vkr+02}). We find that there is
no significant correlation between the pulse intensity prior to, or
after, a radio-off phase. We also find that there is no significant
modulation in the emission during a given radio-on phase. That is, the
variations in pulse intensity are dominated by random
fluctuations. This suggests that the particle flow in the
magnetosphere of PSR~B1931$+$24 remains constant during the radio-on
emission phases, to the limit of our measurement sensitivity and the
intrinsic flux variation of the source.

\begin{figure}
  \begin{center}
    \includegraphics[trim = 2mm 2mm 1mm 5mm, clip, height=8.3cm,width=5.5cm,angle=270]{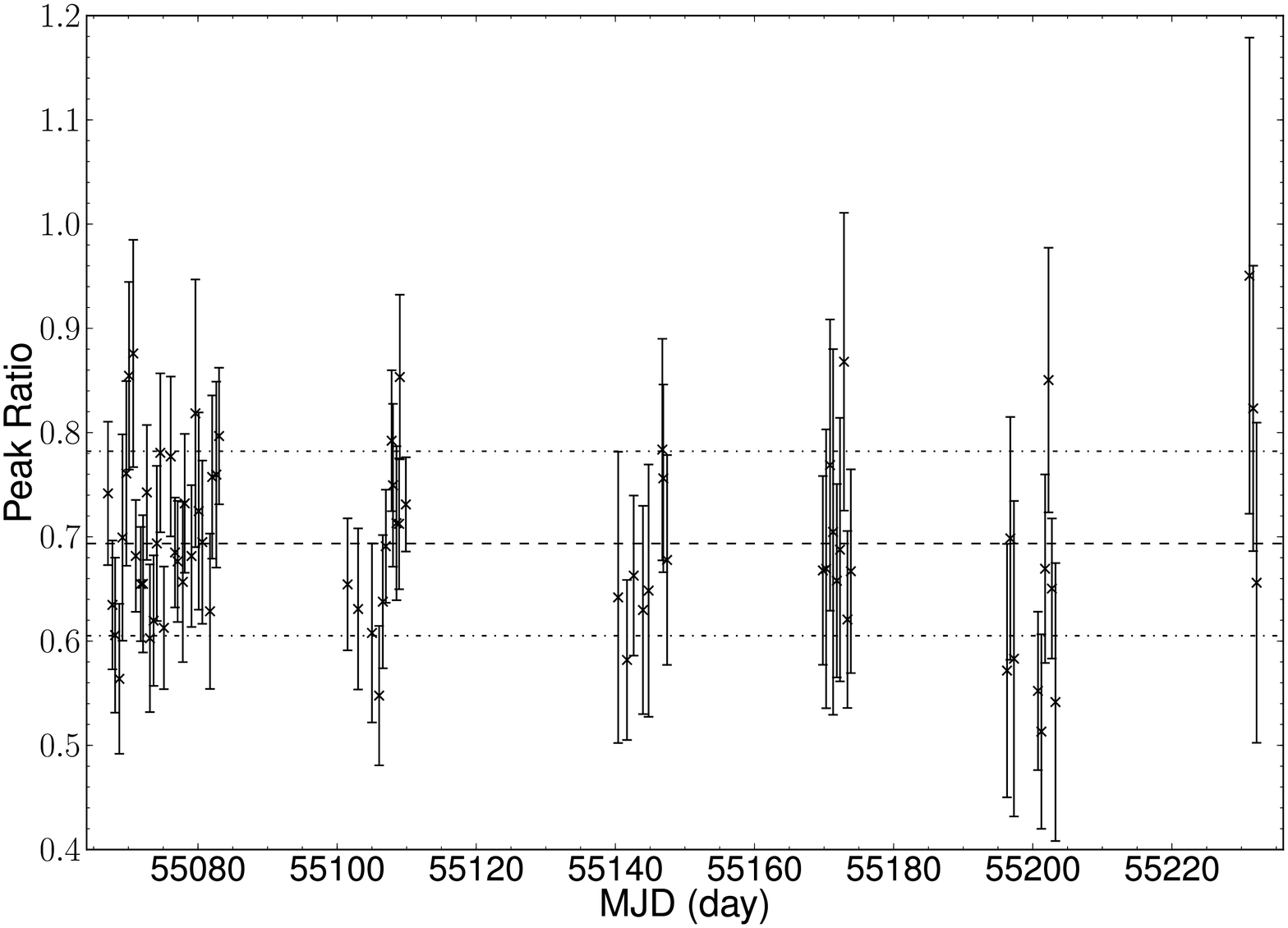}
  \end{center}
 \vspace{-5pt}
\caption{The evolution in the ratio of the first peak to second peak
  intensities of PSR~B1931$+$24, over $\sim165$~d of DFB
  observations. The average and $\pm1\sigma$ values are denoted by the
  (\emph{dashed}) and (\emph{dot-dashed}) lines respectively. The
  fluctuation in the peak ratio values does not appear to be
  significant with respect to the average.}
\label{pkratios}
\end{figure}

To check whether there were any pulse shape changes during the
radio-on phases, as seen in other pulsars exhibiting period derivative
changes \citep{lhk+10}, we examined the variation in the ratio of the
first to second component peak intensities over time. For this
analysis, we aligned DFB profiles with an analytic
template\footnote{This template was produced with the
  {\fontfamily{cmr}\small\selectfont paas} program, which was used to
  fit von-Mises functions to the highest SNR observation. For an
  overview see \hbox{http://psrchive.sourceforge.net/changes/v5.0/}.}
using a $\chi^2$ minimisation method to match the pulse longitudes of
the peak bins. This, in turn, enabled us to quantify the component
peak intensities over time. To obtain a correct representation of the
peak-to-peak ratios, we only performed this analysis on the highest
SNR observations \hbox{($\textrm{SNR}>10$)}. Figure~\ref{pkratios}
shows the distribution of peak ratios. The average peak ratio is
$0.69\pm0.09$ and is found to be consistent with that of the analytic
template (i.e. $\sim0.7$).

In order to determine whether the peak ratio variations were
significant, we again performed an Anderson-Darling test on the
distribution of values, and found no evidence to suggest the data
departs from a normal distribution, assuming Gaussian error bars. As a
consistency check, we also implemented a reduced $\chi^2$ test to
quantify the fluctuations in the pulse shape. For these observations
we obtain $\chi^2 \sim 0.92$, which is consistent with random noise
dominating the profile variation.

\subsection{Periodicity analysis}\label{sec:wwz}
To complement, and improve upon, the results of \cite{klo+06} we have
conducted an extensive analysis on the periodicity of radio emission
from PSR~B1931$+$24, using a much longer data set (i.e. $\sim13$~yr of
data). Here, we use a modified wavelet analysis, i.e. \emph{weighted
  wavelet Z-statistic} (WWZ) analysis \citep{fos96,ysw+12}, to reveal
any periodic variations within the data.

\begin{figure*} 
  \begin{center}
    \includegraphics[trim = 14mm 0mm 7mm 2mm,clip,height=14.5cm,width=10cm,angle=270]{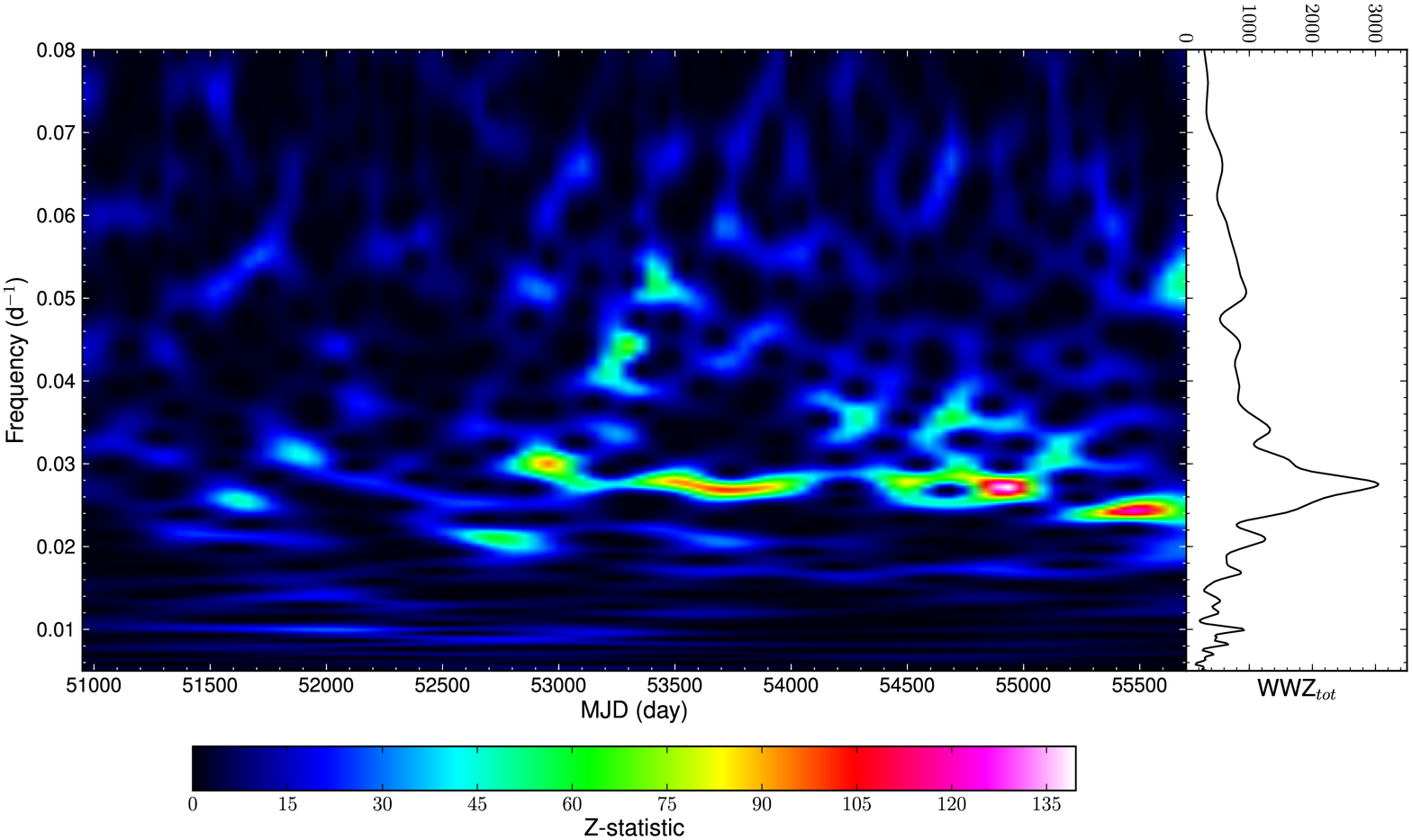}
  \end{center}
\vspace{-7pt}
\caption{WWZ transform of the $\sim13$-yr PSR~B1931$+$24 radio
  emission activity data-set (\emph{left}) and the corresponding
  integrated power spectrum (\emph{right}). The integrated power
  spectrum reaches a maximum at $\sim0.028$~d$^{-1}$
  ($\sim36$~d). Throughout the transform plane the peak frequency
  modulates over time, typically ranging
  \hbox{$\sim0.024-0.032$~d$^{-1}$} ($\sim31-42$~d). The bootstrap
  $5$-$\sigma$ significance level corresponds to
  $Z(\omega,\tau)\approx60$ for this data-set.}
\label{13yr_wwz}
\end{figure*}

For the purpose of our analysis, we chose a wavelet tuning constant
$c=0.001$ so that we could strike a balance between frequency and time
resolution. The resultant WWZ transform of the one-bit time-series
data, showing the evolution of the spectral power at successive epochs
(or \emph{time lags}), is displayed in Fig.~\ref{13yr_wwz}. In the 2-D
transform plot, we note that the peak frequencies, i.e. WWZ frequency
maxima, modulate over time. This, in turn, gives rise to several
features in the integrated spectrum. The broad prominent peak in the
integrated spectrum shows that the WWZ power is preferentially
distributed around $\sim0.031-0.024$~d$^{-1}$ ($\sim32-42$~d), which
corresponds well with the results of \cite{klo+06}. We also note the
presence of other features in the transform plane at
$\sim0.021$~d$^{-1}$ ($\sim48$~d), $\sim0.035$~d$^{-1}$ ($\sim29$~d),
approximately $0.042$~d$^{-1}$ ($\sim24$~d) and $\sim0.052$~d$^{-1}$
($\sim19$~d) over small time intervals. However, they generally have
much lower spectral power which casts doubt on their significance. We
find that the features centred on $\sim0.042$~d$^{-1}$ and
$\sim0.052$~d$^{-1}$ are most likely spurious due to their proximity
to a long, approximately 100-d gap in the data\footnote{Nothing
  meaningful can be obtained for a fluctuation period which is shorter
  than the length of a gap between data points.}. We note that the
better observation sampling towards later times (after
MJD~$\sim52810$) results in better resolution of the intrinsic
variation of the source, that is the fluctuation periods are generally
represented by greater WWZ power towards later times. We also consider
the error estimation of these data, following \cite{ysw+12}, in the
next subsections.

\subsubsection{Confusion limit estimation method}\label{sec:HWHM}
\begin{figure}
  \begin{center}
    \includegraphics[trim = 8mm 3mm 6mm 4mm,clip,height=8.3cm,width=5.5cm,angle=270]{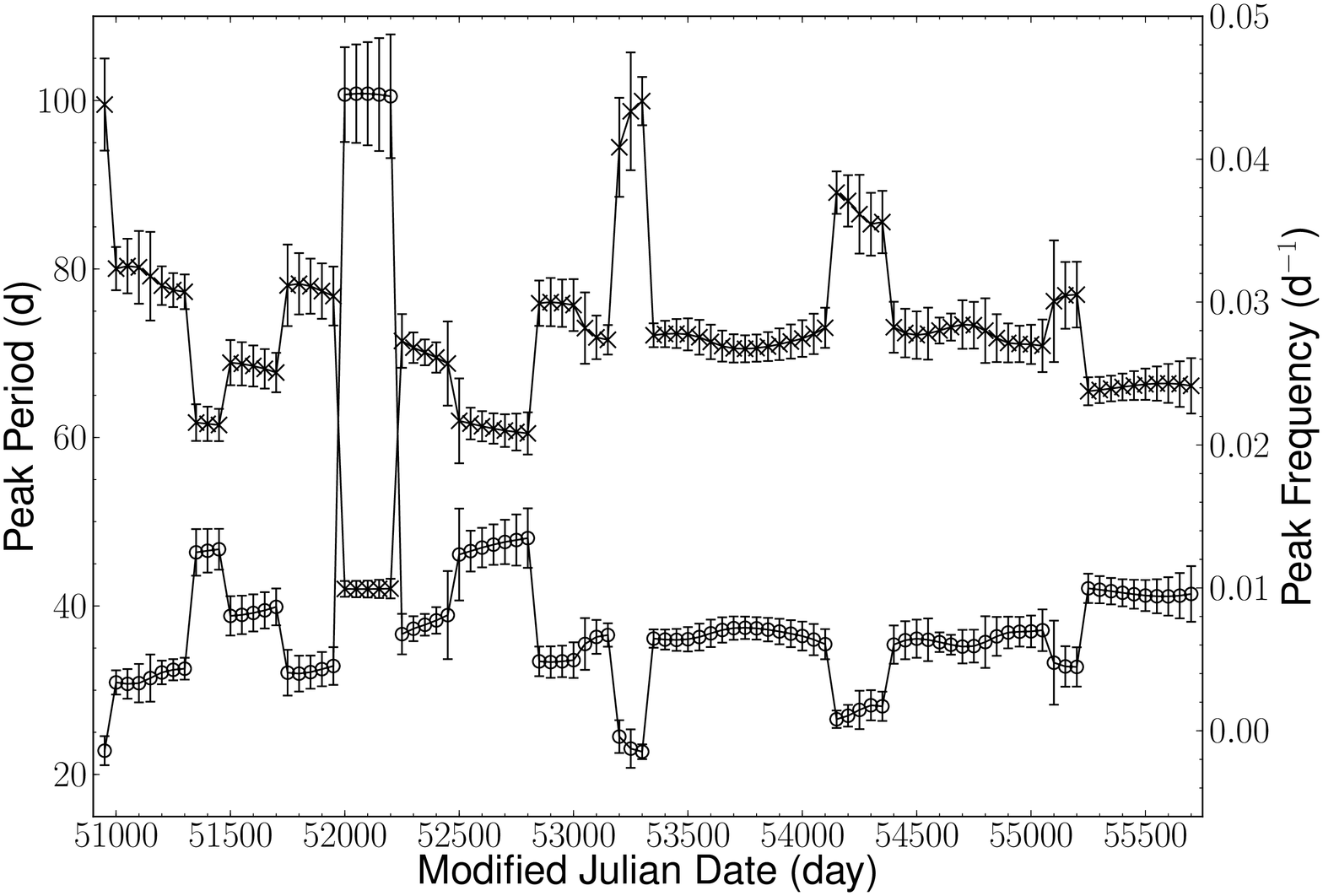}
  \end{center}
\vspace{-6pt}
\caption{Peak WWZ fluctuation frequencies (\emph{crosses}) and periods
  (\emph{open circles}) for the $\sim13$-yr PSR~B1931$+$24 radio
  emission activity data. Error bars are $1$-$\sigma$ values computed
  using the confusion limit estimation method. The average error in
  the fluctuation frequency (or period) is approximately $6~\%$. Note
  that the peak fluctuation period spans $\sim20-50$~d apart from at a
  few epochs where the data sampling is likely to have affected the
  local matching.}
\label{periods}
\end{figure}

To estimate the significance of spectral features, we first
implemented the \emph{confusion limit} method (see \citealt{tmw05,
  ysw+12}). For this analysis, we located the peak fluctuation
frequency (taken to be the frequency bin with the greatest spectral
power) at each epoch and estimated its maximum $1$-$\sigma$
uncertainty as the half-width at half-maximum of the Z-statistic,
$Z(\omega,\,\tau)$. The result of this analysis is shown in
Fig.~\ref{periods}. We note that the peak period spans $\sim20-50$~d
over time, with a few epochs showing periodicities of
$\sim100$~d. However, the features in the wavelet transform attributed
to this longer period are only represented by low WWZ power, and are
over an interval of time which has a number of gaps between successive
epochs ($\mathrm{MJD}\sim51987-52034$, $\sim52040-52077$ and
$\sim52207-52240$). Therefore, we do not believe that the longer
fluctuation period is intrinsic to the source.

We performed another Anderson-Darling test on the peak-fluctuation
period data to determine whether the variations were significant. From
the results of this test, we find that the short-term modulation in
the source's periodicity (see steps in Fig.~\ref{periods}) is not
consistent with random variation. In fact, we find that this
modulation becomes more significant with increasing observation
sampling. As a result, we attribute these short-term fluctuations to
the quasi-periodic nature of the object.

\subsubsection{Data-windowing method}\label{sec:wwzwindow}
As a consistency check, we also used a \emph{data-windowing} method to
compute a measure of the modulation in the peak frequency and period
(see \citealt{ysw+12}). Here, we split the WWZ data into 400-d
segments and, for each of these segments, determined the median
fluctuation frequency and corresponding period (see
Table~\ref{tab:fluc}). These quantities were then compared with the
`noise background' of the WWZ data to determine their
significance. Here, the noise background was estimated by computing
the standard deviation of the transform data, by bootstrapping over
several hundred realisations. Subsequently, we find that the first
four data segments do not have median peak WWZ values greater than the
$5\sigma\approx60$ significance level derived from
bootstrapping. However, the rest of the segments (excluding the data
spanning $\mathrm{MJD}=54150-54550$) have values greater or equal to
this cut-off, which indicate that the periodicities within these data
intervals are intrinsic to the source. We note that the uncertainty in
the peak fluctuation frequency becomes consistently smaller at later
epochs, which results from an increase in resolution due to the
improved data sampling. From this analysis, we obtain median peak
fluctuation frequencies \hbox{$\sim0.035-0.021$~d$^{-1}$}
\hbox{($\sim28-47$~d)}, which are largely consistent with the previous
analysis. Through performing another Anderson-Darling test, we find
that the variations in the long-term periodicity of the source,
i.e. over several years, are not significant. As a result, we infer
that the overall average periodicity in the pulsar radio emission
($38\pm5$~d) is highly stable over long timescales (i.e. years).

\begin{table}
\caption{Results from the data-windowing WWZ error analysis. We quote
the MJD ranges of the data windows analysed, the corresponding
median-peak WWZ values $\mathrm{WWZ}_{\mathrm{max}}$, as well as the
peak fluctuation frequencies $\nu_{\mathrm{fluc}}$ and periods
$P_{\mathrm{fluc}}$. The standard $1$-$\sigma$ uncertainties are
quoted in the parentheses and are in units of the least significant
digit.}
\label{tab:fluc}
\centering
\vspace{4pt}
\begin{tabular}{c  r  r  r}
  \hline
  \hline
   MJD range & WWZ$_{\mathrm{max}}$ & $\nu_{\mathrm{fluc}}$($10^{-2}$~d$^{-1}$) & $P_{\mathrm{fluc}}$(d) \\
  \hline
  50950$-$51350 &  18 (3)  &  3.2 (6)    &  31 (6)  \\
  51350$-$51750 &  30 (10) &  2.5 (3)    &  40 (5)  \\
  51750$-$52150 &  28 (7)  &  3 (1)      &  30 (40) \\
  52150$-$52550 &  23 (4)  &  2.6 (7)    &  40 (30) \\
  52550$-$52950 &  60 (20) &  2.1 (4)    &  47 (7)  \\
  52950$-$53350 &  60 (20) &  3.0 (7)    &  34 (6)  \\
  53350$-$53750 &  90 (10) &  2.75 (4)   &  36 (1)  \\
  53750$-$54150 &  80 (30) &  2.7 (3)    &  37 (3)  \\
  54150$-$54550 &  50 (20) &  3.5 (5)    &  28 (4)  \\
  54550$-$54950 &  80 (30) &  2.80 (5)   &  36 (1)  \\
  54950$-$55350 &  70 (30) &  3.1 (2)    &  32 (3)  \\
  55350$-$55700 & 100 (20) &  2.4205 (1) & 41.3 (2) \\
  \hline
\end{tabular}
\end{table}

\subsubsection{WWZ analysis summary}
The results from the WWZ analysis above show that PSR~B1931$+$24
exhibits quasi-periodicity in its radio emission switching over
timescales of weeks to months. That is, the source displays a
relatively wide range of periodicities ($\sim20-50$~d), when
considering its short-term emission variation. Over longer timescales
(i.e. years), however, the neutron star exhibits a highly stable
average periodicity ($38\pm5$~d) in its radio emission. This implies
that the mechanism governing the source's behaviour is a non-random
systematic process, which is perturbed over timescales of weeks to
months.

\section{Spin evolution}\label{sec:spin}
\subsection{Residual fitting and charge density estimation}\label{sec:resfit}
To gain context on the emission modulation of PSR~B1931$+$24, we
studied its rotational behaviour. We were particularly interested in
determining whether the object experiences any temporal evolution in
the spin-down rates associated with the different phases of emission
which, in turn, may be correlated with alterations to the charge
density distribution of the pulsar magnetosphere (see, e.g.,
\citealt{tim10}).

\begin{figure} 
  \begin{center}
    \includegraphics[trim = 0mm 0mm 0mm 0mm,clip,height=8.2cm,width=6cm,angle=270]{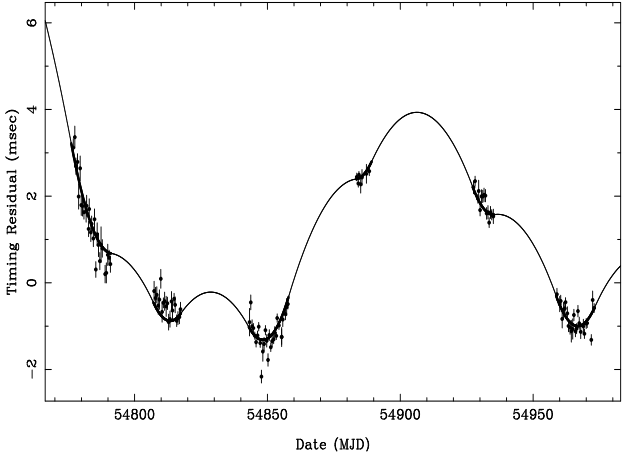}
  \end{center}
\vspace{-5pt}
\caption{Least-squares, weighted fit to timing residuals of
  PSR~B1931$+$24 for observations ranging
  $\mathrm{MJD}\sim54777-54973$. The timing residuals (\emph{filled
    circles}) here are calculated with respect to a fixed $\dot{\nu}$
  ephemeris, while the fitted model (\emph{solid line}) assumes two
  spin-down rates (see Eqn.~\ref{eq:spin_model}).}
\label{fit}
\end{figure}

To quantify the spin-down rate variation in PSR~B1931$+$24, we took a
different approach compared with conventional phase-coherent timing
analysis (see, e.g., \citealt{lpgc96}), due to the emission cessation
in this source. \cite{klo+06} demonstrated the success of using a dual
spin-down rate, least-squares fitting method to derive the
spin-parameters for this pulsar over a short span of data and, hence,
the spin-down rates corresponding to each emission phase. We have
expanded upon this initial analysis by applying the same fitting
procedure to eight well-sampled data sets (including the one presented
in \citealt{klo+06}). The procedure entails minimising the timing
residuals of PSR~B1931$+$24 (calculated using a fixed $\dot{\nu}$
ephemeris) by fitting for the change in rotational phase, period and
period derivative ($\Delta \phi$\footnote{The change in rotational
  phase is the offset required, with respect to an original set of fit
  parameters, to obtain a phase-coherent timing solution due to
  uncertainty in the radio emission phase transition epochs.}, $\Delta
P$ and $\Delta \dot{P}$ respectively) with respect to the separate
emission phases. The values of $\dot{P}_{\mathrm{on},\,\mathrm{off}}$,
and hence $\dot{\nu}_{\mathrm{on},\,\mathrm{off}}$, were derived from
the difference between the average spin-down rate
($\dot{P}_{\mathrm{av}}$) and $\Delta \dot{P}$ associated with these
modes. The total $\Delta \phi$, $\Delta P$ and $\Delta \dot{P}$, which
were required to obtain a phase-coherent solution, were modelled via:
\begin{equation}
  t_{\mathrm{res}} = \Delta \phi \cdot P + \frac{\Delta P}{P} \cdot (t - t_{\mathrm{i}}) + \frac{\Delta \dot{P}_{\mathrm{i}}}{2 P} \cdot (t - t_{\mathrm{i}})^2\, ,
\label{eq:spin_model}
\end{equation}
where $t_{\mathrm{res}}$ and $t_{\mathrm{i}}$ are the residual
($t_{\mathrm{obs}}-t_{\mathrm{pred}}$) and reference times
respectively (here $t_{\mathrm{pred}}$ represents the time-of-arrival
predicted by a single spin-down rate model), and $\Delta
\dot{P}_{\mathrm{i}}$ is the change in spin-down rate in each emission
phase (i.e. $\pm\Delta\dot{P}$). An example of this fitting process is
shown in Fig.~\ref{fit}. The rotational behaviour of the pulsar is
clearly characterised well by the model.

We note that the total contribution of
$\dot{\nu}_{\mathrm{on},\,\mathrm{off}}$ to the average is determined
from the total time spent in a given emission phase, which is
inherently dependent on the transition times between
phases. Typically, we can only constrain transition times to within
about a day which, in turn, imparts systematic errors into this
analysis. To propagate the effects of such errors we performed a
Monte-Carlo simulation on the data. Here we performed $10^5$ fits, for
each data interval, using transition times which are randomly
distributed between the last radio-on and first radio-off phase
observations. As a result, we were able to obtain a distribution of
radio-on and -off spin-down rates for each data interval, from which
we determined estimates for their averages and associated
uncertainties (see Table~\ref{tab:fits}). We note that the overall
average values, that is
\hbox{$\langle\dot{\nu}_{\mathrm{on}}\rangle=-16 \pm 1 \times
  10^{-15}$~s$^{-2}$} and
\hbox{$\langle\dot{\nu}_{\mathrm{off}}\rangle=-10.8 \pm 0.4 \times
  10^{-15}$~s$^{-2}$}, are found to be consistent with \cite{klo+06}.

Following \cite{klo+06}, we assume that the spin-down in the radio-off
states is governed by pure magnetic dipole radiation. Whereas, we
assume that the spin-down in the radio-on states is supplemented by an
additional torque due to an outflowing plasma (i.e. a particle wind)
and its associated return current. Using this model, we can estimate
the plasma charge density attributed to this wind component, to an
order of magnitude, via (see \citealt{klo+06}, and references therein,
for more details):
\begin{equation}
  \rho_{\mathrm{plasma}} \approx
  \frac{3I\,\Delta\dot{\nu}}{R_{\mathrm{pc}}^4\,B_{\mathrm{s}}}\,,
\label{eq:rho_plasma}
\end{equation}
where $I$ is the moment of inertia of the object (taken to be
$10^{45}$~g~cm$^2$), $\Delta\dot{\nu}$ is the change in spin-down rate
of the pulsar between emission states
($\dot{\nu}_{\mathrm{on}}-\dot{\nu}_{\mathrm{off}}$), $R_{\mathrm{pc}}
= \sqrt{2\pi R^3 \nu/c}$ is the polar cap radius of the pulsar (where
the pulsar radius $R$ is taken to be $10^6$~cm; see, e.g.,
\citealt{lk05}) and the surface magnetic field strength (for the
simplest case of an orthogonal rotator; see, e.g., \citealt{jac62}) is
\begin{equation}
  B_{\mathrm{s}}\sim3.2\times10^{19}\,\sqrt{\left(\frac{-\dot{\nu}_{\mathrm{off}}}{1\,\rm{s}^{-2}}\right)\!\left(\frac{\nu}{1\,\rm{s}^{-1}}\right)^{\!-3}}\,\rm{Gauss}\,.
\label{eq:Bs}
\end{equation}
By substituting $R_{\mathrm{pc}}$ and $B_{\mathrm{s}}$ into
Eqn.~\ref{eq:rho_plasma}\footnote{Note that an additional factor of
  $10^7/c$ is used to convert between CGS units
  ($\mathrm{statC}\,\mathrm{cm}^{-3}$) and SI units
  ($\mathrm{C}\,\mathrm{m}^{-3}$), so as to be consistent with the
  literature (see also Eqn.~\ref{eq:rho_GJ}).},
$\rho_{\mathrm{plasma}}$ was directly calculated for each data
interval via:
\begin{equation}
  \rho_{\mathrm{plasma}}\sim 7.1\times10^5 \!\left(\frac{\Delta\dot{\nu}}{1\,\rm{s}^{-2}}\right)\!\!\left(\frac{1\,\rm{s}^{-1}}{\nu}\right)^{\!\!1/2}\!\!\left(\frac{1\,\rm{s}^{-2}}{\dot{\nu}_{\mathrm{off}}}\right)^{\!\!1/2}\!\rm{C}\,\rm{m}^{-3}.
\label{eq:rho_plasma_sub}
\end{equation}

These quantities were also compared with the Goldreich-Julian charge
density of the object, which was estimated, to an order of magnitude,
by (see, e.g., \citealt{lk05}):
\begin{equation}
  \rho_{_{\mathrm{GJ}}} = \frac{B_{\mathrm{s}} \,\nu}{c}\sim
  3.6\times10^5 \left(\frac{\dot{\nu}_{\mathrm{off}}}{1\,\rm{s}^{-2}}\right)^{\!\!1/2}\!\!\left(\frac{1\,\rm{s}^{-1}}{\nu}\right)^{\!\!1/2}\mathrm{C}\,\mathrm{m}^{-3}.
\label{eq:rho_GJ}
\end{equation}
Using the average values for
$\dot{\nu}_{\mathrm{off}}=-10.8\pm0.4\times10^{-15}$~s$^{-2}$ and
$\nu=1.228965\pm0.000001$~Hz, obtained from the fitting process, we
estimate $\rho_{_{\mathrm{GJ}}}$ to be approximately $0.033
\,\rm{C}\,\rm{m}^{-3}$. Table~\ref{tab:fits} shows the results of this
analysis. We note that different values are measured for
$\dot{\nu}_{\mathrm{on}}$ and $\dot{\nu}_{\mathrm{off}}$ over
neighbouring data intervals. However, given the uncertainties in these
parameters, we consider the different spin-down rates for each
emission phase to be consistent with the overall averages of
$\dot{\nu}_{\mathrm{on}}$ and $\dot{\nu}_{\mathrm{off}}$. This is
supported by the results of an Anderson-Darling test, from which we
conclude that the distributions of $\dot{\nu}_{\mathrm{on}}$ and
$\dot{\nu}_{\mathrm{off}}$ follow Gaussian statistics. As a result, we
do not consider the variations in $\rho_{\mathrm{plasma}}$ with
respect to $\rho_{_{\mathrm{GJ}}}$ to be significant.

\begin{table*}
\caption{Summary of the results from the residual fitting
  analysis. The epoch of the spin parameters for each fit is denoted
  by $\mathrm{Pepoch}$, $T$ is the length of each fit interval and
  $\dot{\nu}_{\mathrm{on}}$, $\dot{\nu}_{\mathrm{off}}$ are the
  radio-on and radio-off spin-down rates respectively. The change in
  spin-down rate
  (i.e. $\dot{\nu}_{\mathrm{on}}-\dot{\nu}_{\mathrm{off}}$) is
  represented by $\Delta \dot{\nu}$. Definitions for
  $\rho_{_{\mathrm{GJ}}}$ and $\rho_{\mathrm{plasma}}$ can be found in
  the text. Standard $1$-$\sigma$ errors are provided in the
  parentheses, after the parameters (where possible), and represent
  the least significant digit.}
\label{tab:fits}
\begin{center}
\small
\begin{tabular}{l  r  r  r  r  r  r  r}
  \hline
  \hline
  $\mathrm{Pepoch}$ & $T$ (d) & $\dot{\nu}_{\mathrm{on}}$ ($10^{-15}$~s$^{-2}$) & $\dot{\nu}_{\mathrm{off}}$ ($10^{-15}$~s$^{-2}$) & $\Delta \dot{\nu}/\dot{\nu}_{\mathrm{av}}$ ($\%$) & $\rho_{\mathrm{plasma}}$ ($10^{-2}$~Cm$^{-3}$) & $\rho_{\mathrm{plasma}}/\rho_{_{\mathrm{GJ}}}$ ($\%$)\\
  \hline
  51869.7 & 107.5   & -15.2 (2)  & -11.13 (9) & 36.6 (6)  & 2.5 & 73  \\
  52842.3$^*$ & 157.5 & -16.0 (2)  & -10.78 (7) & 48.4 (7) & 3.2 & 96 \\
  53782.6 & 97.7   & -17.6 (5)  & -10.4 (2)  & 69 (2)   & 4.5  & 134  \\
  53987.1 & 193.4  & -19 (1) & -10.2 (2)  & 83 (5)   & 5.4  &  160  \\
  54340.7 & 326.8  & -16.2 (4)  & -10.7 (1)  & 51 (1)   & 3.4 & 102 \\
  54676.6 & 133.0  & -17.1 (8)  & -10.3 (3)  & 66 (4)   & 4.3  & 130  \\
  54875.0 & 196.0  & -15.8 (2)  & -11.07 (8) & 42.7 (6) & 2.9 & 87  \\
  55198.2 & 262.2  & -14.3 (1)  & -11.40 (3) & 25.4 (2) & 1.7 & 52 \\
  \hline
\end{tabular}
\end{center}
\begin{flushleft}
\scriptsize{\hspace{13.5mm}$^{*}\mathrm{Pepoch}$ of data used to derive spin-parameters in \cite{klo+06}.}
\end{flushleft}
\end{table*}

\subsection{Measurement of the braking index}\label{sec:BI}
In the radio-off phase, the rotational slow-down of the pulsar
$\dot{\nu}_{\mathrm{off}}$ is thought to be indicative of the torque
produced by magneto-dipole radiation \citep{klo+06}. Consequently, the
rate of change in the spin-down rate in the radio-off phase
$\ddot{\nu}_{\mathrm{off}}$ should provide a measure of the braking
index associated with this radiation (e.g. \citealt{tm77})
\begin{equation}
n=\frac{\nu\ddot{\nu}_{\mathrm{off}}}{\dot{\nu}^2}\,,
\label{eq:BI}
\end{equation}
which is not contaminated by the additional electromagnetic torque of
the plasma current flow; that is, assuming that the additional torque
completely disappears in the radio-off phase. This is particularly
important as measurement of this parameter can offer significant
information about how the pulsar undergoes its energy loss and,
ultimately, provide us with insight into the electrodynamics (see,
e.g., \citealt{xq01,wxg03,rud05,con07}).

\begin{figure}
  \begin{center}
    \includegraphics[trim = 2mm 0mm 1mm 5mm,clip,height=8.3cm,width=5.5cm,angle=270]{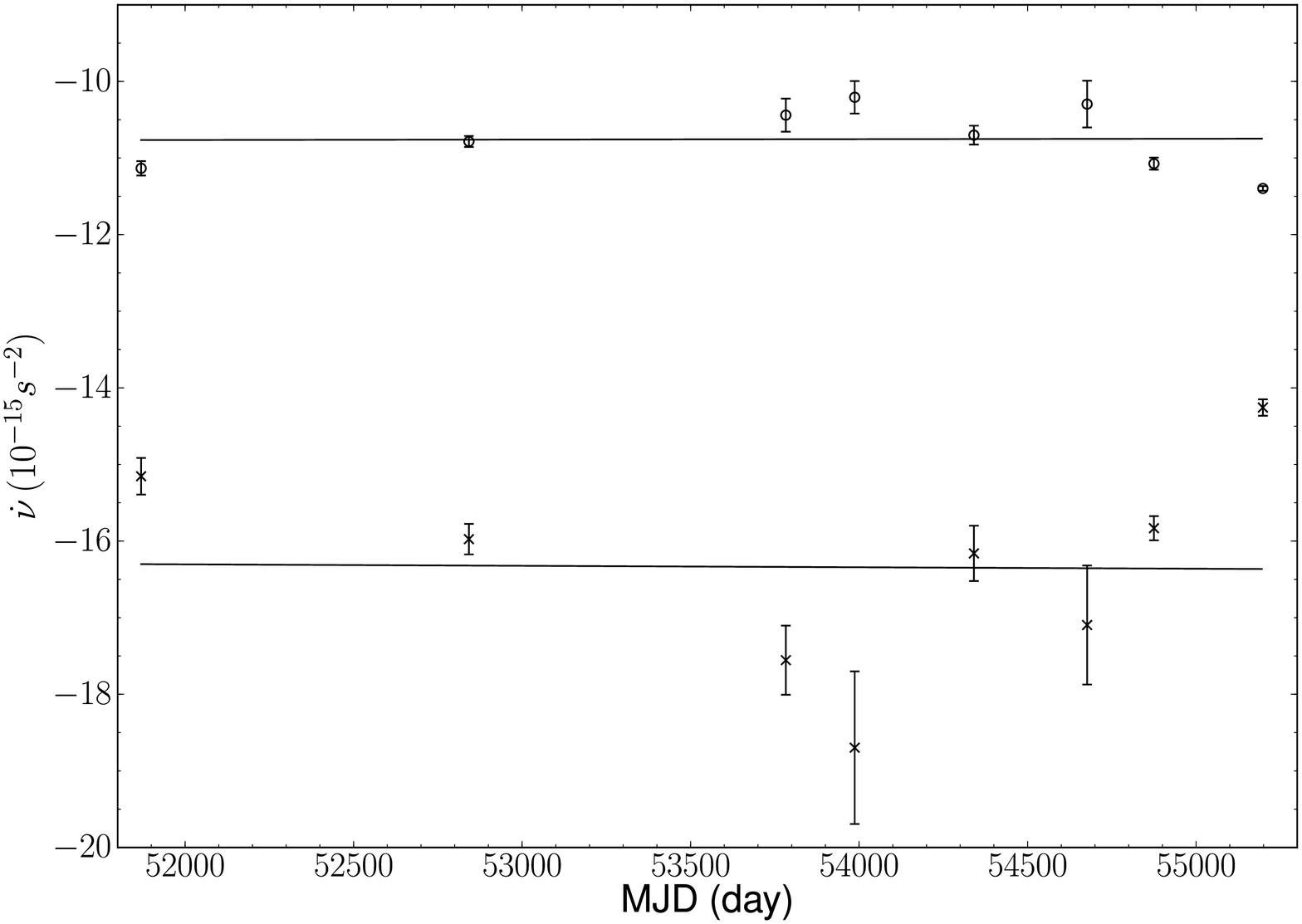}
  \end{center}
\vspace{-5pt}
\caption{Linear regression fits of $\dot{\nu}_{\mathrm{off}}$
  (\emph{top~trace}) and $\dot{\nu}_{\mathrm{on}}$
  (\emph{bottom~trace}) for PSR~B1931$+$24. Note the anti-correlation
  in the data, which affects the quality of the fits (see text for
  details).}
\label{nuddot}
\end{figure}

Here, we fitted the derived values for $\dot{\nu}_{\mathrm{on}}$ and
$\dot{\nu}_{\mathrm{off}}$ for a number of epochs to determine if
significant second order variations could be measured
(Fig.~\ref{nuddot}). Unfortunately, however, we could not obtain a
significant value for $\ddot{\nu}_{\mathrm{off}}$ using data even with
the greatest observation density (Table~\ref{tab:BI}, see also
$\S$\ref{sec:discuss} for a discussion on this). As a result, we could
not obtain an accurate measure of the braking index of the source.

\begin{table}
\caption{The results of the linear regression analysis of the
  spin-down rate data, with least significant uncertainties in the
  parentheses. The length of each fitted data-set is denoted by
  $T$. The correlation coefficient and two-sided probability value for
  each fit are given by $R$ and $p$ respectively. The rate of change
  in the spin-down rate is represented by $\ddot{\nu}_{\mathrm{off}}$,
  with the associated $1$-$\sigma$ error quoted in the parentheses (in
  units of the least significant digit).}
\label{tab:BI}
\centering
\vspace{4pt}
\begin{tabular}{r  r  r  r  r }
  \hline
  \hline
  $T$ (d) & $R$ & $p$ & $\ddot{\nu}_{\mathrm{off}}$ ($10^{-19}$~s$^{-3}$) \\
  \hline
  3328.5 &  0.01 & 0.97 & 0.05 (156) \\
  2355.9 & -0.42 & 0.35 & -2 (2)     \\
  1415.5 & -0.78 & 0.06 & -7 (3)     \\
  \hline
\end{tabular}
\end{table}

Consequently, we sought to determine a value for braking index using
another method. Using the timing software developed by \cite{wje11},
we also modelled the variation in spin-down rate in PSR~B1931$+$24, by
fitting changes in rotational phase to the times-of-arrival (using the
first three terms of Eqn.~7 in \citealt{wje11}). These changes in
rotational phase are analogous to glitches, that is they are discrete
changes in spin-down rate, although there are no jumps in spin
frequency. In our model, each emission phase has a unique spin-down
rate (i.e. there is only one value for $\dot{\nu}_{\mathrm{on}}$ and
$\dot{\nu}_{\mathrm{off}}$), which we assume is acquired directly
after (or before) the last (or first) radio-on observation of each
active emission phase. Through fitting the observed times-of-arrival
for changes in spin-down rate only, we find that this method produces
results consistent with Table~\ref{tab:fits}. However, when the
emission phase transition times are also taken as a fit parameter we
obtain very poor fits. We applied maximum and minimum time constraints
to the `glitch epochs' but, due to the absence of data bounding a
given transition into (or out of) a radio-on phase, the fitting
process could not converge; a global solution could not be achieved
because times-of-arrival are limited to the radio-on phase data.

\subsection{Long-term evolution in spin-down rate}\label{sec:nudotevol}
Using the results from the residual fitting in $\S$\ref{sec:resfit},
we were unable to determine a significant value for
$\ddot{\nu}_{\mathrm{off}}$ or $n$ for the pulsar. This is most likely
due to the transition time errors, between emission phases, which
introduce an anti-correlation between $\dot{\nu}_{\mathrm{on}}$ and
$\dot{\nu}_{\mathrm{off}}$ (a linear fit to these data obtains a
correlation coefficient $R=-0.9$ and two-sided p-value $p=0.0003$; see
also Fig.~\ref{nuddot}). That is, systematic over- and
under-estimation of the transition times, which are not fitted for,
cause $\dot{\nu}_{\mathrm{on}}$ and $\dot{\nu}_{\mathrm{off}}$ to
become anti-correlated which, in turn, increases the noise
contribution to $\ddot{\nu}_{\mathrm{off}}$.

\begin{figure} 
  \begin{center}
    \includegraphics[trim = 2mm 1mm 1mm 3mm,clip,height=8.3cm,width=5.5cm,angle=270]{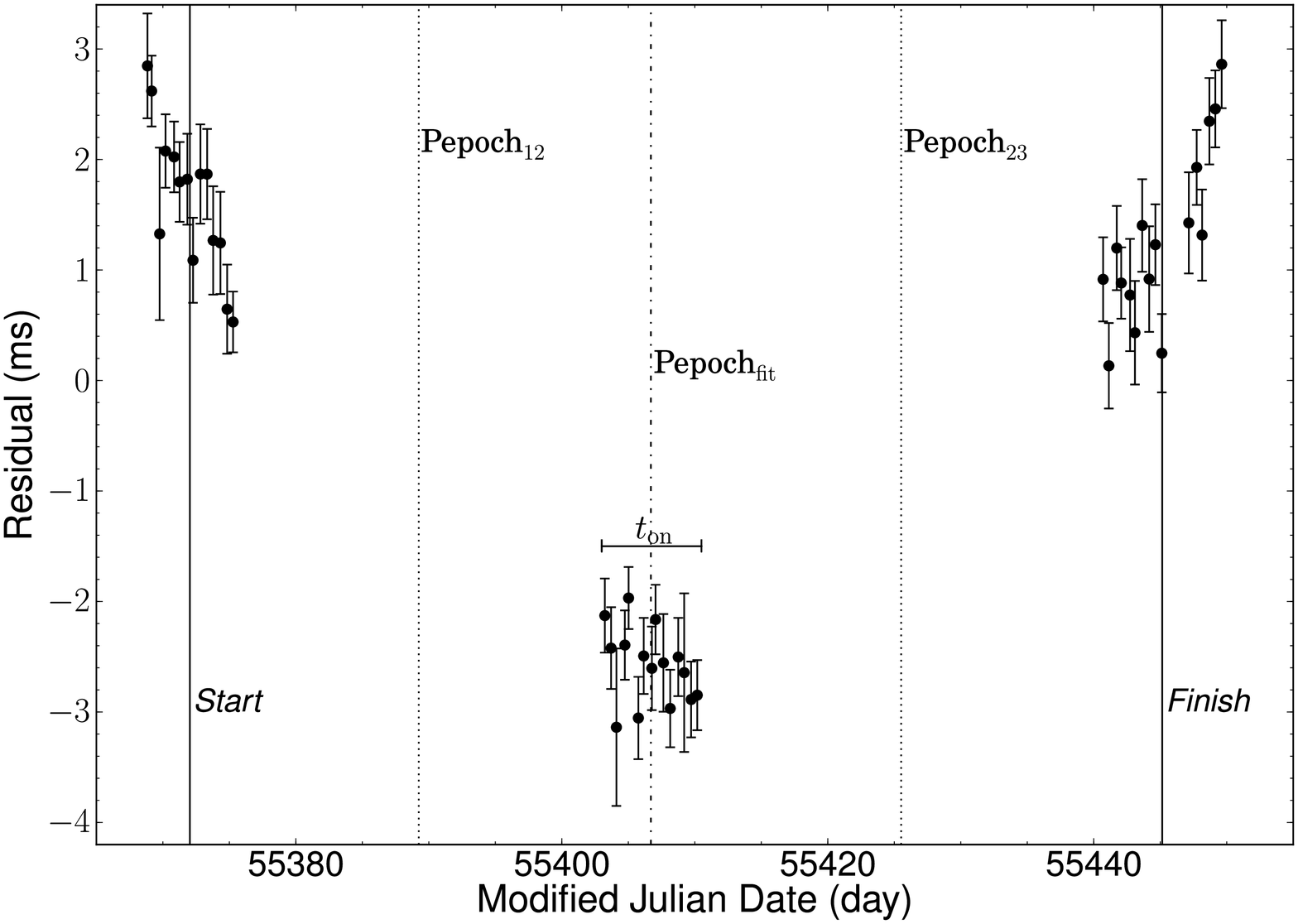}
  \end{center}
\vspace{-3pt}
\caption{Timing residuals of PSR~B1931$+$24, over approximately $80$~d
  of data, depicting the $\Delta \nu$ fitting process. The epochs
  where $\nu$ was calculated are represented by
  $\textrm{Pepoch}_{\,12}$ and $\textrm{Pepoch}_{\,23}$ for the first
  and second pair of radio-on phases respectively. The epoch at which
  $\Delta \nu$, hence $\Delta \dot{\nu}$, was measured is denoted by
  $\textrm{Pepoch}_{\,\mathrm{fit}}$. The bounding regions of the fits
  are shown by the `start' and `finish' lines.}
\label{deltanu}
\end{figure}

To complement this analysis, therefore, we decided to examine the
long-term evolution in the spin-down rates over the entire $\sim13$-yr
data-set. For this analysis, we used timing measurements of
PSR~B1931$+$24 to estimate the contribution of
$\dot{\nu}_{\mathrm{on}}$ to the rotational frequency $\nu$ over
neighbouring radio-on phases. We assumed that the change in rotational
frequency $\Delta \nu$, resulting from variation between
$\dot{\nu}_{\mathrm{on}}$ and $\dot{\nu}_{\mathrm{off}}$, can be
obtained from fitting the residuals of three successive radio-on
phases in separate pairs (see Fig.~\ref{deltanu}). In order to
separate the effects of the different spin-down rates in these fits,
we first form residuals by using a fixed $\dot{\nu}$, set to be the
average $\dot{\nu}_{\mathrm{off}}$ (using the average determined from
residual fitting,
i.e. $\langle\dot{\nu}_{\mathrm{off}}\rangle=-10.8\pm0.4\times10^{-15}$~s$^{-2}$). We
fitted the timing residuals of the latter and first halves of a given
pair of radio-on phases, using
{\fontfamily{cmr}\footnotesize\selectfont
  PSRTIME}\footnote{http://www.jb.man.ac.uk/pulsar/observing/progs/psrtime.html},
to estimate a value for the rotational frequency which is governed by
the radio-off spin-down rate. We assumed that the contribution of
$\dot{\nu}_{\mathrm{on}}$ to these residuals was negligible due to the
short length of time the radio-on phases cover (w.r.t. the radio-off
phases). By obtaining values for $\nu$, for the first and second pair
of radio-on phases, we were able to estimate the total change in
rotational frequency due to the difference in spin-down rate $\Delta
\dot{\nu}$ over the central radio-on phase. Given that
$\dot{\nu}_{\mathrm{on}}$ should only affect the residuals over this
period, we obtain the relation
\begin{equation}
  \Delta \nu = \nu_{23} - \nu_{12} = \Delta \dot{\nu} \times t_{\mathrm{on}}\,,
\label{eq:deltanu}
\end{equation}
where $\Delta \dot{\nu}$ represents
$\dot{\nu}_{\mathrm{on}}-\dot{\nu}_{\mathrm{off}}$ and
$t_{\mathrm{on}}$ is the total time spent in the central radio-on
phase. The spin-frequencies for the first and second fit intervals are
$\nu_{12}$ and $\nu_{23}$ respectively, as shown in
Fig.~\ref{deltanu}.

To increase the number of measurements, we analysed the data using a
stride-fitting method similar to that used by \cite{lhk+10}; we
analysed groups of radio-on phase data in steps of single emission
phases. We only included data windows which had three well defined
radio-on phases i.e. containing gaps no greater than $10$~d between
observations\footnote{The minimum total time spent in a consecutive
radio-on and -off phase is $10$~d. A gap between residuals of this
timescale would cause a large uncertainty in the contribution of
$\dot{\nu}$ to $\Delta \nu$, due to possible unmodelled rotational
behaviour.}, with uncertainties in the emission phase transitions less
than $4$~d (less than $2$~d for the longest data-set; see
Table~\ref{tab:deltanu}). We corrected for the long-term spin-down
behaviour of the source by subtracting the average spin-down rate from
the frequency data using the {\fontfamily{cmr}\small\selectfont vfit}
program\footnote{http://www.jb.man.ac.uk/pulsar/observing/progs/vfit.html}.

\begin{figure}
  \begin{center}
    \includegraphics[trim = 0mm 3mm 1mm 5mm,clip,height=8.3cm,width=5.5cm,angle=270]{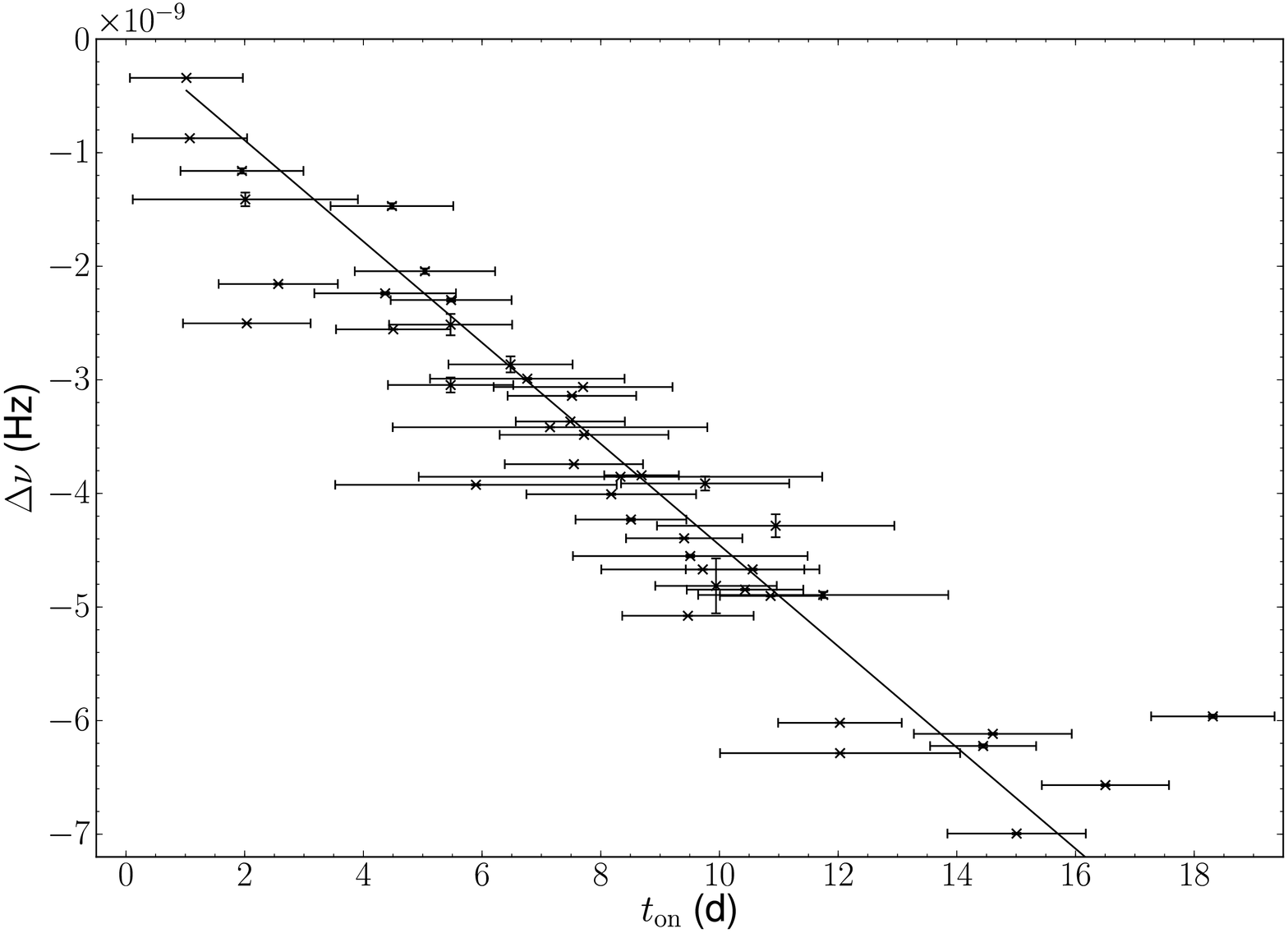}
  \end{center}
\vspace{-7pt}
\caption{A linear fit to the change in rotational frequency ($\Delta
  \nu$) against radio-on phase length ($t_{\mathrm{on}}$) for
  approximately 13~yr of observations of PSR~B1931$+$24.}
\label{deltanufit}
\end{figure}

Noting that Eqn.~\ref{eq:deltanu} is analogous to the equation for a
straight-line, with a gradient equal to $\Delta \dot{\nu}$ and zero
intercept, we were able to fit $\Delta \nu$ vs $t_{\mathrm{on}}$
across numerous data intervals, as shown in Fig.~\ref{deltanufit} and
Table~\ref{tab:deltanu}. We find that these data are highly linearly
correlated, which suggests that $\dot{\nu}_{\mathrm{on}}$ and
$\dot{\nu}_{\mathrm{off}}$ are well defined. We find that there is no
evidence for variation in the spin-down rates associated with the
different phases of emission within our measurement sensitivity.

\begin{table}
\caption{The linear fit parameters for the long-term spin-down rate
  study of PSR~B1931$+$24. The MJD mid-point of each data-set used is
  denoted by $\mathrm{MJD}_{\mathrm{mid}}$. The total length of each
  data-set is represented by $T$ and $N$ is the number of data points
  used in each fit. The change in spin-down rate obtained from the
  fitting procedure is denoted by $\Delta \dot{\nu}$ and
  $\dot{\nu}_{\mathrm{on}}$ is the corresponding radio-on spin-down
  rate. Standard ($1$-$\sigma$) errors for $\Delta \dot{\nu}$ and
  $\dot{\nu}_{\mathrm{on}}$ are provided in parentheses after the
  parameters, and represent the least significant digit.}
\label{tab:deltanu}
\centering
\vspace{4pt}
\begin{tabular}{r  r  r  r  r  r}
  \hline
  \hline
  $\mathrm{MJD}_{\mathrm{mid}}$ & $T$ (d) & $N$ & $\Delta \dot{\nu}$ ($10^{-15}$~s$^{-2}$) & $\dot{\nu}_{\mathrm{on}}$ ($10^{-15}$~s$^{-2}$)\\
  \hline
  51646.0 & 1426 & 8  & -5.6 (8) & -16 (1)   \\
  53143.4 & 1451 & 9  & -4 (1)   & -16 (1)   \\
  54602.8 & 1467 & 29 & -5.2 (3) & -16.0 (5) \\
  53356.2 & 4688 & 45 & -5.2 (2) & -16.0 (4) \\
  \hline
\end{tabular}
\end{table}

\section{Discussion}\label{sec:discuss}
We have shown that PSR~B1931$+$24 exhibits regulated emission
modulation; that is, to the limit of our measurements it maintains a
constant ADC. To improve on these results, even more frequent sampling
would be required. This would only be feasible with a large dedicated
telescope, or new facilities such as LOFAR and the SKA\footnote{Phase
1 science operations with the SKA are projected to commence in 2020,
with full operation of the telescope proposed in 2024. See also
http://www.skatelescope.org/about/project/ for more details.} which
will have multi-beam capabilities and, hence, greater ability to
dedicate observing time to individual sources. If we were to observe
this source over hourly timescales, rather than daily, we would
dramatically increase our chances of observing the pulsar transition
between emission modes. As a result, we would possibly be able to
uncover correlated changes in emission following (or preceding)
radio-off phases which might give us insight into what causes the
magnetospheric configuration to change so dramatically.

We also characterised the modulation timescales of the radio emission
in PSR~B1931$+$24. We find that the source exhibits a periodic
modulation timescale of approximately 38~d on average. There do appear
to be variations around this basic periodicity, but it remains
remarkably stable over many years. This degree of stability provides a
challenge for models of this process as it is significantly longer
than the expected dynamic and plasma timescales ($\sim$~ms; e.g.
\citealt{rs75}).

Using a residual fitting method, we find that the variations in
$\dot{\nu}_{\mathrm{on}}$ and $\dot{\nu}_{\mathrm{off}}$, throughout
the $\sim13$-yr data set, are consistent with the measurement
uncertainties (see $\S$~\ref{sec:resfit}). This result is supported by
the long-term analysis (see $\S$~\ref{sec:nudotevol}), which provides
additional evidence to suggest that we do not observe any significant
changes in the spin-down rates attributed to each emission mode; that
is, the pulsar appears to retain a constant $\Delta\dot{\nu}= 48\pm
2\%$ between emission phases. This implies that the apparent changes
in plasma flow in the radio-on phase are also not significant
($\langle\rho_{\mathrm{plasma}}\rangle =4\pm1\times10^{-2}$~Cm$^{-3}$;
see Table~\ref{tab:fits}), thus indicating a surprisingly high degree
of stability in the bi-modal system.

We note, however, that there is an unavoidable systematic effect in
the short-timescale fitting procedure. More specifically, the errors
in the transition times between emission phases introduced uncertainty
into the calculation of $\dot{\nu}_{\mathrm{on}}$ and
$\dot{\nu}_{\mathrm{off}}$. This is highlighted by the results in
Fig.~\ref{nuddot}, where the interdependency between the two spin
parameters is made evident. We see that these quantities are in fact
almost perfectly anti-correlated. In our model, the transition times
between phases of emission are not taken as a fit
parameter. Therefore, observations spaced farther apart between two
emission modes will have a higher uncertainty in the transition time
compared with two spaced closer together. Furthermore, if the amount
of time spent in a radio-on phase is overestimated, then the amount of
time in a consecutive radio-off phase will be underestimated
accordingly. Given that radio-on phases are much shorter than
radio-off phases on average, it is evident that the effect on the
error of $\dot{\nu}_{\mathrm{on}}$ will be greater than that of
$\dot{\nu}_{\mathrm{off}}$. This naturally explains the observed
anti-correlation between the two parameters and the larger error bars
in $\dot{\nu}_{\mathrm{on}}$. Therefore, it is possible that we did
not see any significant variation in the spin-down rates due to this
systematic effect. To overcome this problem, we again would require
greater observation cadence to provide better constraints on the
transition times.

We also attempted to obtain a value for the braking index of
PSR~B1931$+$24 from $\ddot{\nu}_{\mathrm{off}}$, in the hope of
determining information about the energy loss of the object. Using
Eqn.~\ref{eq:BI}, we predicted the $\ddot{\nu}$ that would be required
to obtain a braking index $n=3$. Assuming the average parameters for
$\nu$ and $\dot{\nu}_{\mathrm{off}}$ from residual fitting, we would
expect $\ddot{\nu} = 2.85 \pm 0.05 \times 10^{-28}$~s~$^{-3}$
(i.e. $\sim10^9$ less than the upper limit derived from fitting
separate residual data-sets). Given the level of timing noise in this
pulsar, and the $\ddot{\nu}$ quoted in
\cite{hlk10}\footnote{\cite{hlk10} obtain $\ddot{\nu}= 7.0 \pm 0.1
  \times 10^{-25}$~s~$^{-3}$ for a time span of $\sim13.7$~yr, which
  is a factor of $10^6$ less than the upper limit derived from this
  work. As they apply a global fit to the timing data, this lower
  limit is corrupted by the spin-down rate in the radio-on phase.}, it
seems very improbable that any increase in data density would allow us
to obtain a value for $\ddot{\nu}$ as low as that predicted from
theory. The problem lies in decoupling the long-term contribution of
$\dot{\nu}_{\mathrm{on}}$ from the timing residuals; that is, the
value for $\ddot{\nu}$ from global fitting will always be contaminated
by the modulation of $\dot{\nu}$. In the case of local fitting
($\dot{\nu}_{\mathrm{on},\,\mathrm{off}}$), there will probably always
be an uncertainty in the transition times which, again, means that it
will be difficult to decouple $\dot{\nu}_{\mathrm{on},\,\mathrm{off}}$
from $\dot{\nu}_{\mathrm{av}}$.

With the next generation of telescopes (e.g. LOFAR, FAST, ASKAP,
MeerKAT and the SKA) coming online in the near
future\footnote{Although LOFAR is already collecting data, the
  complete array is yet to be fully built.}, we anticipate a dramatic
increase in the number of known intermittent
pulsars\footnote{Approximately 900 pulsars are predicted to be
  discovered with LOFAR alone \citep{vs10}, many of which will likely
  be transient objects.}. Coupled with the continued observation of
known intermittent sources (e.g. PSRs~B0823$+$26, J1832$+$0029,
J1841$-$0500 and B1931$+$24), characterisation of these objects should
lead to an improved understanding of the general emission and
rotational properties of pulsars which, ultimately, should facilitate
the development of more realistic radio emission and magnetospheric
models. These observations should also provide further insight into
how the different `types' of nulling pulsars, i.e. normal nulling
pulsars, RRATs and longer-term intermittent pulsars, are related.

\section{Conclusions}\label{sec:conc}
In this work, we have confirmed and expanded substantially upon the
findings of \cite{klo+06}, which has enabled us to provide further
evidence for the magnetospheric-state switching scenario
(e.g. \citealt{bmsh82,lhk+10,tim10}). In order to determine how these
events are triggered (e.g. via `circumpulsar asteroids';
\citealt{cs08}), however, it is clear that dedicated infra-red and
high-energy observations of this source are required.

If the spin-down rates attributed to each emission mode, and ADC, for
this pulsar are truly constant over time, as suggested by our
observations, this would be quite remarkable; as of yet, there is no
clear reason to suggest why a pulsar should retain a memory of its
previous magnetospheric state i.e. particle flow
(c.f. \citealt{lst12a}). This raises a couple of important questions:
1) Is there charge or matter transfer in the magnetosphere which leads
to these separate regulated states and, if so, how would this occur?
2) Why does the pulsar consistently assume the same rotational and
emission characteristics? These questions are likely only to be
answered based on future observations of this source and others like
it, which may as yet be undiscovered. We note, however, that the rate
at which new intermittent sources sources are discovered should
increase significantly with the use of next generation telescopes.
Therefore, we may only need to wait a few years before significant
breakthrough is made in magnetospheric modelling and our understanding
of emission cessation in pulsars.

\section{Acknowledgements}
We thank D.~J.~Champion and M.~D~Gray for useful discussion which has
contributed to this paper. We are grateful to C.~Jordan, and the
several telescope operators at Jodrell Bank, for obtaining the
majority of the data used in this work. NJY acknowledges support from
the National Research Foundation (NRF). The Nan\c{c}ay Radio
Observatory is operated by the Paris Observatory, associated with the
French Centre National de la Reserche Scientifique (CNRS).

\bibliographystyle{mn2e} 
\bibliography{journals,psrrefs,njy_modrefs}
\end{document}